\documentclass[useAMS,usenatbib]{mn2e}

\usepackage{graphicx}

\usepackage{latexsym}

\usepackage{amsmath}

\usepackage{subfigure}

\usepackage{wasysym}

\title[Discs evolution and dispersal]{Protoplanetary disc evolution and
    dispersal: the implications of X-ray photoevaportion}

\author[J. E. Owen, B. Ercolano \&
    C. J. Clarke]{James E. Owen$^1$\thanks{E-mail: jo276@ast.cam.ac.uk},
    Barbara Ercolano$^{2}$ \& Cathie J. Clarke$^{1}$ \\ $^1$Institute of
    Astronomy, Madingley Road, Cambridge, CB3 0HA, UK\\ $^2$School of
    Physics, University of Exeter, Stocker Road, Exeter EX4 4QL}

\begin{document}

\pagerange{\pageref{firstpage}--\pageref{lastpage}} \pubyear{2002}

\newcommand{\dd}{\textrm{d}}

\newcommand{\rin}{R_\textrm{\tiny{in}}}

\newcommand{\OO}{\mathcal{O}}

\newcommand{\msun}{M$_\odot$}

\newcommand{\msunyr}{M$_\odot$~yr$^{-1}$}

\maketitle

\label{firstpage}

\def\mnras{MNRAS}

\def\apj{ApJ}

\def\aap{A\&A}

\def\apjl{ApJL}

\def\apjs{ApJS}

\def\araa{ARA\&A}

\begin{abstract}
We explore the role of  X-ray photoevaporation  in the evolution and
dispersal of viscously evolving T-Tauri discs. We show that the X-ray photoevaporation wind rates scale
linearly with X-ray luminosity, such that the observed range of X-ray
luminosities for solar-type T-Tauri stars (10$^{28}$-10$^{31}$
erg s$^{-1}$) gives rise to vigorous disc winds with rates of
order 10$^{-10}$-10$^{-7}$~M$_{\odot}$ yr$^{-1}$. These mass-loss rates  are
comparable to typically observed T-Tauri accretion rates, immediately
demonstrating the relevance of X-ray photoevaporation to disc evolution. 
We use the wind solutions from radiation-hydrodynamic models, coupled to a
viscous evolution model to construct a population synthesis model so that we may 
study the physical properties of evolving discs and so-called
`transition discs'. Current observations of disc lifetimes and
accretion rates can be matched by our model assuming a viscosity
parameter $\alpha = 2.5 \times 10^{-3}$. 

Our models confirm that X-rays play a dominant role in the evolution
and dispersal of protoplanetary discs giving rise to the observed 
diverse population of inner hole `transition' sources which include
those with massive outer discs, those with gas in their inner holes and
those with detectable accretion signatures. 
To help understand the
nature of observed transition discs we present a diagnostic diagram
based on accretion rates versus inner hole sizes that demonstrate
that, contrary to recent claims, many of the observed accreting
and non accreting transition discs can easily be explained by
X-ray photoevaporation. However, we draw attention to a smaller but still significant, population of strongly accreting
($\sim$10$^{-8}$~\msunyr) transition discs with large inner
holes ($>$ 20~AU) that lie outside the predicted X-ray photoevaporation region, suggesting a
different origin for their inner holes. 

Finally, we confirm the conjecture of Drake et al. (2009), that
    accretion is suppressed by the X-rays through `photoevaporation
    starved accretion' and predict this effect can give rise to a
    negative correlation between X-ray luminosity and accretion rate,
    as reported in the Orion data. We also demonstrate that our model
    can replicate the observed difference in X-ray properties between
    accreting and non-accreting T-Tauri stars.  

\end{abstract}


\section{Introduction}

Protoplanetary discs are a natural outcome of low mass star formation,
providing a reservoir of material from which the star itself continues
    to accrete and from
which planets may later form. The accretion history of a newly formed
    star, the evolution of its
disc and the formation of a planetary system are all intimately
linked and affected by irradiation from the central star. 
A lot of attention has recently been paid to
the final dispersal of protoplanetary discs as this sets the time-scale over which planets
may form.
Observationally, the study of the dust content of these discs through the analysis of their continuum
spectral energy distributions (SEDs), in the infra-red (IR), has
    enormously advanced over the past few decades. These observations
    have indicated discs evolve in a way that suggest `standard' viscous evolution
    (Hartmann et al. 1998)
    for the first few Myrs (e.g. Haisch et al. 2001b), but then
    rapidly evolve from a disc-bearing (primordial) to
disc-less status (e.g. Luhman et al. 2010; Muzerolle et al. 2010). This rapid clearing of the
    inner disc indicated from IR observations has been
    supplemented with complementary observations in the sub-millimetre that show most
    non-accreting stars (WTTs) are devoid of emission out to $\sim500$~AU
    (Duvert et al. 2000). This suggests the removal of material
    material close to the star ($<1$ AU) is correlated with the
    removal of material at larger radius, ($>10$ AU); and is hence, associated with the entire destruction of the protoplanetary disc. 

Furthermore, there have been several observations of
    discs thought to be caught in the act of clearing (e.g Calvet et
    al. 2002; Najita et al. 2007); these `transition' discs show a deficit in emission at
    near-IR (NIR) wavelengths compared to a primordial optically thick disc (one that
    is optically thick all the way to the dust destruction radius), while
    returning to the emission levels expected from an optically thick
    disc at longer wavelengths. Currently, there is no clear
    consensus as to what constitutes a transition disc and both
    accreting (e.g. Hughes et al. 2009; Espaillat et al. 2010) and
    non-accreting (e.g. Cieza et al. 2010; Mer{\'{\i}}n et al. 2010) objects have
    been discovered.  

The frequency of these inner-hole sources is approximately 10 to 20\% of
the total disc population for young solar-type stars in nearby star forming regions (e.g Strom et al. 1989; Skrutskie et al. 1990; Luhman et al. 2010). If one assumes that these
inner hole objects are indeed `transition' discs (i.e. discs that are
caught in the act of dispersal) and that the gas and dust evolution
go hand-in-hand,
then a 10\% frequency of transition discs suggest a dispersal
mechanism that operates over a tenth of the optically thick
disc lifetime (Kenyon \& Hartmann 1995).  Moreover the colours of
    such discs imply that they clear from the inside out (Ercolano,
    Clarke \& Hall, 2010). 

The observations described above have encouraged theorists to propose
dispersal mechanism that operate 
over two separate time-scales: a first time-scale of a few million years
which allows discs to 
 remain optically thick at IR wavelengths, followed by a brief dispersal phase that 
leaves the discs optically thin with colours consistent with those of bare photospheres or debris
discs. The proposed mechanisms include planet formation itself
(e.g. Armitage \& Hansen 1999), grain growth (Dullemond \& Dominik 
2005), photophoresis (Krauss et al 2007), MRI-driven winds (Suzuki \& Inutsuka 2009), and photoevaporation
due to extreme ultraviolet radiation -EUV- (Clarke et al  2001; Alexander et al 2006a,b; Richling \& Yorke 2000), X-ray radiation (Alexander et al. 2004; Ercolano et
al 2008b, 2009; Gorti \& Hollenbach 2009, Owen et al. 2010) and far ultraviolet radiation
-FUV- (Gorti \& Hollenbach 2009; Gorti et al. 2009). However there is still no consensus over which mechanism (or which combination)
may dominate.

The recent developments of X-ray photoevaporation (XPE) models for
T-Tauri discs have yielded encouraging results, suggesting 
that this may be the dominant player in disc dispersal. In particular,
we have shown that X-ray photoevaporation rates exceed the EUV
photoevaporation rates by two orders of magnitude (Ercolano et al
2009) and can easily reproduce the two time-scale 
behaviour suggested by the observations (Owen et al 2010). This model
does not suffer from the uncertainties in the heating rates that
plague the FUV model, due to the unknown abundance of PAHs in discs and
the very uncertain photoelectric yields (see discussion in Ercolano \&
Owen 2010). Furthermore, Ercolano \& Clarke (2010) discussed the role of metallicity in disc dispersal, and found that photoevaporation predicts shorter disc lifetimes at lower metallicity, in contrast to planet formation which would predict a very strong negative correlation.  
Yasui et al (2009) have presented observation that favour a positive correlation between disc lifetimes and metallicity and hence which favour XPE; however, it is too early to draw any definitive conclusions about the role metallicity plays in disc dispersal. Finally, the slightly  blue-shifted forbidden line
spectrum of discs seen almost face on (e.g. Hartigan et al 1995;
Pascucci et al 2009) is well reproduced by the XPE model for all of
the low ionisation and atomic diagnostics considered (Ercolano \&
Owen, 2010), an improvement over the EUV photoevaporation model which
cannot reproduce the OI 6300\AA~ line intensities (e.g. Font et al 2004).

More recent observations have focused on understanding the nature of
some known transition discs (TDs) by investigating their gas content and accretion
properties. Evidence for accretion and gas inside the dust inner hole
has sometimes been used to argue against
photoevaporation being the agent of inner-hole clearing  
(e.g. Espaillat et al 2010). This is however a misconception, perhaps
based on earlier EUV-based photoevaporation models (Alexander et
al. 2006b). In this work we show that the XPE model of Owen et al
(2010) clearly predicts a transition phase where there is a detectable
quantity of accreting gas inside the dust inner hole and is indeed
consistent with the accretion versus inner hole size properties of
most observed TDs. 

In this paper we use radiation-hydrodynamics simulations to develop a X-ray photoevaporation model which is then used to
construct a synthetic disc population to compare with available
observed disc statistics. We discuss the X-ray luminosities of TTs in Section~2, while Section~3 provides a description of the X-ray photoevaporation model. In Section~4 we discuss a viscously evolving photoevaporating disc model and the construction of our disc population. In Section~5 we present the results from our population synthesis model  while our final conclusions are given in
Section~6.

\section{X-ray Luminosity Function}

The X-ray luminosity ($L_X$), is a crucial input for the X-ray
    photoevaporation model, where we take the X-ray luminosity to span the range $0.1-10$keV. As discussed in Section~3, the choice of
    $L_X$ fully describes the mass-loss rate due to photoevaporation
    in a given system.  

 The input X-ray spectrum is identical to the spectrum used in
    previous work (Ercolano et al. 2008b; Ercolano et al. 2009; Owen et al. 2010; Ercolano \&
    Owen 2010). It is a synthetic spectrum generated by the plasma
    code of Kashyap \& Drake (2000) and fits to {\it Chandra} spectra
    of T Tauri stars (e.g. Maggio et al. 2007). It includes an
    extreme ultraviolet component - EUV - (13.6-100~eV), soft X-ray
    component (0.1-1~keV) and a hard X-ray component
    ($>1$~keV). Ercolano et al. (2009) studied the effect of
    attenuating this X-ray spectrum with neutral columns. They found that at columns of order
    $10^{21}$~cm$^{-2}$ the soft X-ray component was screened out and
    the photoevaporation rates dropped significantly. However, for
    columns $\le 10^{20}$~cm$^{-2}$  the photoevaporation rates
    remained unaffected. Furthermore, Ercolano \& Owen (2010) showed
    that the X-ray wind itself is optically thick to EUV photons: thus
    when considering photoevaporation it is only the strength of
    the soft X-rays that is relevant. We assume that the soft X-rays
    are able to reach the disc surface as is consistent with the
    explanation of the OI 6300\AA~ line emission  presented in Ercolano \& Owen (2010). Therefore, we adopt the unattenuated
    spectrum shown in Owen et al. (2010) and expect our results to
    generalise to cases of moderate screening. We will discuss the
    implications of this assumption further in Section~5.

It has been known
    for some time that there is a large scatter in the X-ray luminosity of
    T-Tauri stars for a given stellar mass (and bolometric luminosity). Therefore, we have used the data
for the Taurus cluster (G\"{u}del et al. 2007) and the Orion Nebular
    cluster (Preibisch et al. 2005) to build cumulative X-ray luminosity functions \footnote{Where the X-ray luminosity is taken to be in the range 0.3-10keV for the Taurus sample and 0.5-8keV for the Orion sample, consistent with our definition which includes both the soft and hard component.} for the Orion sample functions (XLFs)
for all
pre-main sequence stars (including both CTTs and WTTs) in the 0.5 to
    1~M$_{\odot}$ range.  The data for these two regions shows good agreement at low luminosity, but they differ at high
luminosities, in the sense that the Orion data contains a higher
    fraction of sources with high $L_X$. The difference is most probably {\it not} intrinsic to
the X-ray properties of these two regions, but due to the different
treatment of strong flares in the two samples. Strong flares were excluded
in the Taurus data, in contrast to the Orion sample where luminosities
were averaged over the whole observational period. Strong flares are relatively
rare and Albacete Colombo et al. (2007) found that non-flaring sources have a median $KT=2.1\pm0.3$KeV, compared to flaring sources with $KT=3.8$KeV. Therefore, due to their higher X-ray temperature flares tend to emit most
of their radiation in the hard X-ray region (as shown by observations
    of objects in a `flare' state compared to in a `quiescent' state,
    e.g. Imanishi et al. 2001), where the thermal impact
is low, as discussed above. For this reason, the Taurus sample is
the most appropriate for use in the photoevaporation model, since
    it should better approximate the quiescent and therefore softer X-ray luminosity,
    which as discussed above is the relevant input for calculations
    of the photoevaporation rate. Therefore, we adopt the Taurus XLF for the
remainder of this work.

We assume that the X-ray luminosity function remains
    invariant throughout the stars' pre-main sequence evolution. While it
    is known that the median of the stellar X-ray luminosity function
    does in fact decrease with age due to stellar spin down
    (e.g. Hemplemann et al. 1995; G\"{u}del et al. 1997; G\"{u}del 2004), the
    evolution of the    X-ray luminosity         during the disc
    dispersal phase is much smaller (see Figure 41, G\"{u}del 2004). Certainly, any evolution of the X-ray luminosity
    for ages up to several tens of Myrs is smaller than
    the observed spread in X-ray luminosities at $\sim
    1$Myr. Furthermore, there is
    observational evidence that the X-ray luminosities of CTTs and
    WTTs are significantly different, with WTTs in general being  more
    luminous than CTTs (e.g. Neuhauser et al 1995; Stelzer \& Neuhauser 2001; Flaccomio et
    al. 2003; Preibisch et al. 2005). This has lead to a discussion of
    X-ray emission being `disturbed' by accretion, in terms of either X-ray absorption in accretion
    columns (Gregory et al. 2007) or confinement of the X-ray
    producing corona in accreting systems (Preibisch et
    al. 2005). Recently, Drake et al. (2009) suggested that X-rays may
    modulate accretion through
    photoevaporation (something they called `photoevaporation starved
    accretion'). Such a scenario may be able to account for the difference
    in the observed X-ray luminosities of CTTs and WTTs since more
    luminous stars will lose their disc's first.

In order to assess  the effect of `photoevaporation starved
    accretion' in explaining the X-ray observations we adopt here the
    null stance; the X-ray luminosity of an individual young stellar
    object (YSO) remains constant in time as our models evolve from
    CTTs to WTTs
    (i.e. we assume that the X-ray luminosity function is fixed).       

\begin{figure}
\centering
\includegraphics[width=\columnwidth]{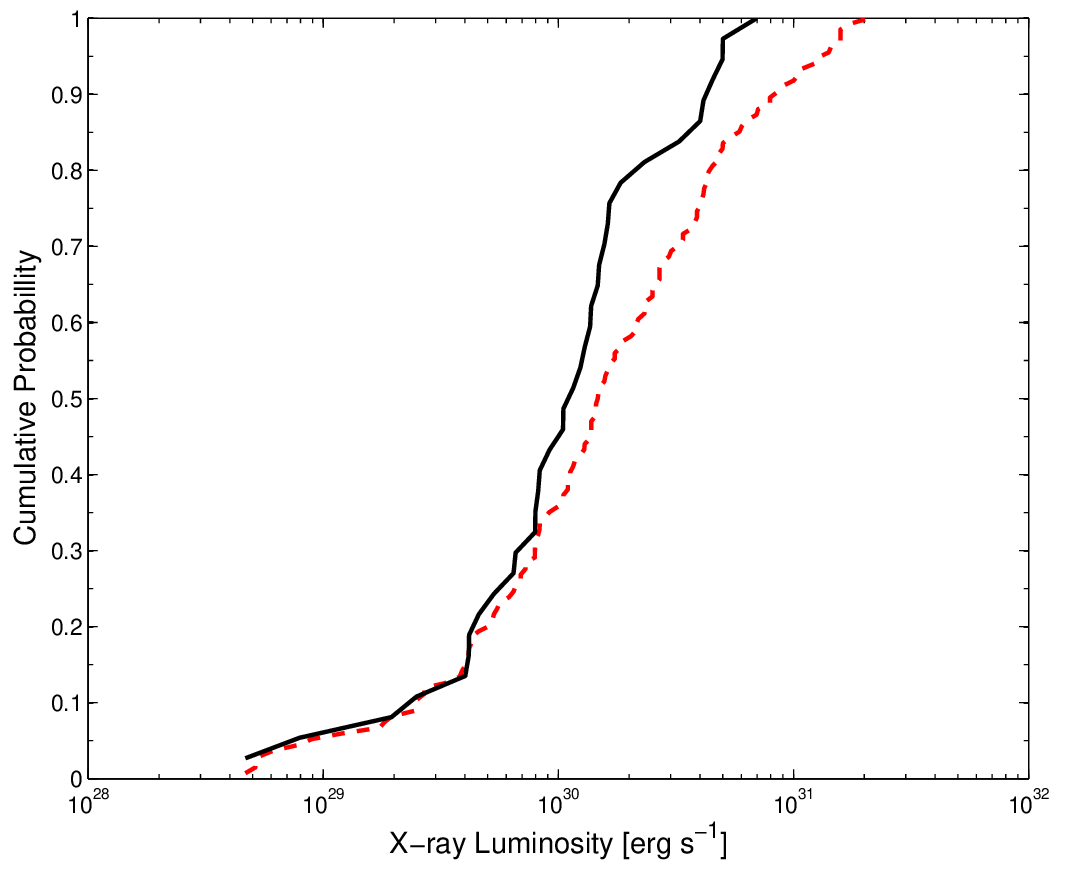}
\caption{Cumulative X-ray luminosity functions
    for the Taurus (black solid line) and Orion (red dashed line) clusters for solar type
    stars with mass in the range [0.5,1.0]~M$_\odot$, including both WTTs and
    CTTs.}\label{fig:xlf}
\end{figure} 

\section{X-ray Photoevaporation}\label{sec:theory}

While the first self-consistent numerical simulations performed by Owen et al. (2010) were a significant step forward in 
understanding X-ray photoevaporation, the models were calculated for
    only one value of the X-ray luminosity ($2\times 10^{30}$erg
    s$^{-1}$). A variety of underlying disc models were
    considered, which showed that X-ray photoevaporation is fairly
    insensitive to the details of the X-ray `dark' region, due to
    the fact that the sonic surface is located at least several flow scale
    heights from the X-ray `dark'/`bright' transition. However, the large
    observed spread in X-ray luminosities means that the dependence of photoevaporation on the X-ray luminosity needs to be considered. 

Before we discuss the results of a detailed numerical investigation of
    parameter space,
    we can use simple fluid mechanics to predict the variation of
    photoevaporation rates with X-ray luminosity. Namely, any
    axi-symmetric inviscid steady state flow must satisfy the
    following conditions: conservation of mass; conservation of
    specific angular momentum; hydrostatic equilibrium perpendicular
    to the streamlines; and the conservation
    of a Bernoulli type potential\footnote{While the conservation of
    the Bernoulli potential does not formally exist for non-barotropic
    flows, it is easy to verify that for the X-ray winds studied here
    (i.e. unconfined and thermally driven), one can find a suitable,
    well defined (single valued) function $P(\rho)$ along each
    streamline that allows us to construct an energy integral
    equivalent to the Bernoulli potential for a baratropic flow with
    the identical $P(\rho)$. By uniqueness, this Bernoulli-type energy
    integral must be a conserved quantity along the streamline in the
    X-ray heated flow.} along each streamline.  As discussed in
    Ercolano \& Owen (2010) the velocity and temperature in a thermally
    driven wind is intimately linked the value of the effective
    potential, and unlikely to be greatly affected by changes in the
    X-ray luminosity. Furthermore, as shown in Owen et al. (2010) the
    gas temperature can be roughly described in terms of the ionization parameter ($\xi=L_X/nr^2$) alone.
Thus if we consider any X-ray heated wind (primordial or a disc with an
    inner hole) and vary the X-ray luminosity, spatially fixing the
    temperature and velocity requires that the ionization parameter
    remains constant. This means that the density will scale {\it
    globally} in the flow as $n\propto L_X$. It is then a trivial matter to show that the new wind solution we have constructed, with a different X-ray luminosity will still satisfy the required conditions described previously. Hence, we can use the conservation of mass to show that the X-ray photoevaporation model predicts:
\begin{equation}
\dot{M}_w\propto L_X
\end{equation} 
for all X-ray flows including both primordial discs and disc with
    inner holes. Also, since the only change is to the global density,
    the normalised mass-loss profiles should also remain approximately
    independent of the X-ray luminosity. Furthermore, Owen et
    al. (2010) also showed that there was very little variation in
    mass-loss rate with inner hole radius.  

\subsection{Radiation-Hydrodynamic Models}\label{hydro}

We have performed radiation-hydrodynamic models to study the form of
the mass-loss, under the assumption that it is
driven by X-ray photoevaporation (XPE). 
We consider XPE from discs around solar-type stars for both primordial
(optically thick discs extending into the dust destruction radius) and discs with
inner-holes (in both gas and dust). 
We run primordial disc models at six equal logarithmic intervals
between $\log L_X= [28.3, 30.8]$. We further calculate four inner
hole models at three different X-ray luminosities $\log L_X =
{28.8,29.8,30.8}$ (in addition to the eight models already computed by 
Owen et al. 2010 at $\log L_X=30.3$), at radii of $\sim 5,10,20,70$~AU.
We start from an initial density structure of a protoplanetary disc
surrounding a 0.7M$_\odot$ star with $T_{\textrm{eff}}$=4000 K and
$R_*$=2.5$R_\odot$, taken from the set of
D'Alessio et al. (2004). 
We adopt the following elemental abundances, given as number
    densities with respect to hydrogen: He/H$=0.1$,
    C/H$=1.4\times 10^{-4}$, N/H$=8.32\times 10^{-5}$,
    O/H$=3.2\times 10^{-4}$, Ne/H$=1.2\times 10^{-4}$,
    Mg/H$=1.1\times 10^{-6}$, Si/H$=1.7\times 10^{-6}$,
    S/H$=2.8\times 10^{-5}$. These are solar abundances
    (Asplund, Grevesse \& Sauval, 2005) depleted according to Savage
    \& Sembach (1996).  
We use the {\sc zeus2D} code (Stone \& Norman 1992) to calculate
the hydrodynamical evolution of the disc, where the gas temperature
due to X-ray irradiation is parametrised as a function of ionisation
parameter using the 3D radiative transfer code {\sc mocassin}
(Ercolano et al 2003, 2005, 2008). The temperature of gas that is not
heated by the X-rays is fixed to the dust temperature and the transition
point between the X-ray bright and the X-ray dark region occurs at a
column of 10$^{22}$ cm$^{-2}$. Such an approximation obviously
    results in a temperature and density discontinuity at this point;
    however, since this transition occurs well below the sonic surface (where
    the mass-loss rate is effectively set) it does not affect the
    resulting mass-loss rate. The distribution evolves to a steady
state, with a bound X-ray dark disc and a thermally driven, X-ray
bright, transonic photoevaporative wind. The numerical methods employed
here and briefly described above are similar to those of Owen et al. (2010) and we refer the reader to that article for a more detailed
description of  the model setup.

\subsubsection{Wind Rates \& Streamline Topology}

The numerical models described above were used to calibrate the
    analytical relations derived at the beginning of Section~3, which we can use to build a synthetic disc population. 
The main results are summarised as follows:

\begin{enumerate}

\item As expected, for a star of a given mass the total mass loss rate scales
almost  linearly with the X-ray luminosity. This result applies to
both primordial and inner hole sources. Specifically,
\begin{equation}\label{eqn:mdotLx}
\dot{M}_w =6.4\times10^{-9}A\left( \frac{L_X}{10^{30} \textrm{erg s}^{-1}}\right)^{1.14}\textrm{\msunyr}
\end{equation}
where $A$ is a constant taking the value unity for primordial discs and
    0.75 for discs with inner holes.
Figure~\ref{fig:mdotLx} shows the mass loss rates calculated by our grid of
radiation-hydrodynamics models for primordial discs. Equation~\ref{eqn:mdotLx} is
also a good fit for inner-hole sources, irrespective of hole size, for
the cases studied here (inner hole radii between 5 and 70~AU).
\begin{figure}
\centering
\includegraphics[width=\columnwidth]{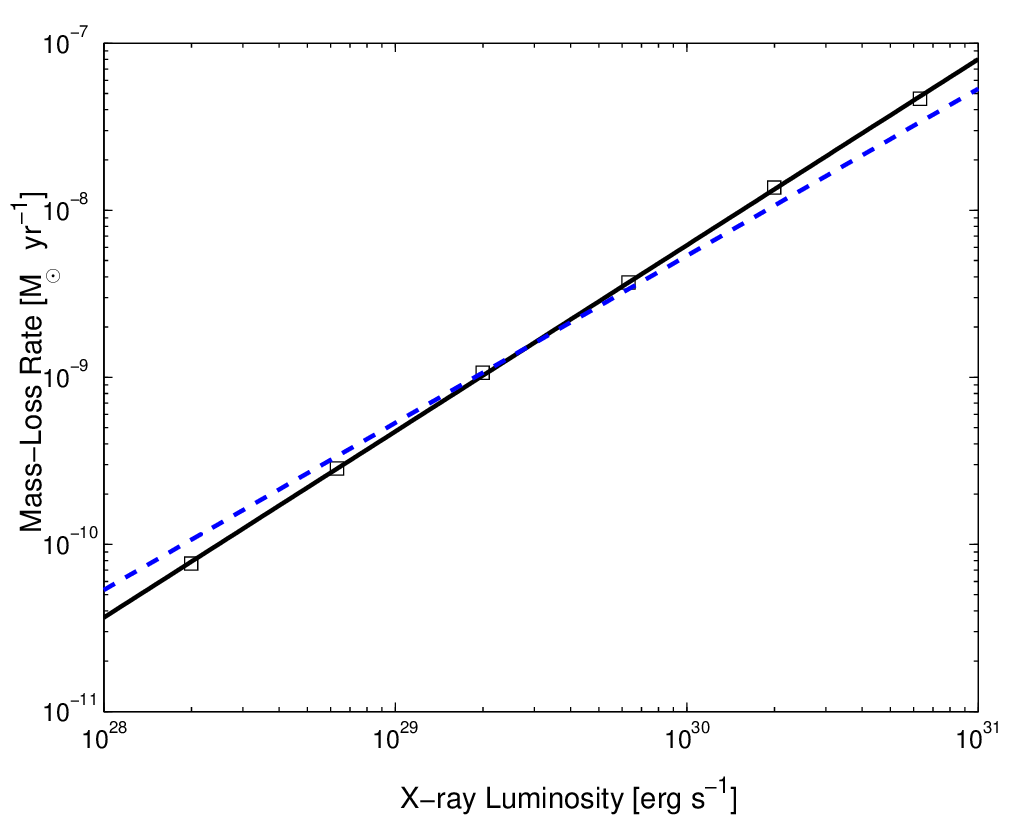}
\caption{Total mass-loss rates as a function of X-ray
    luminosity. Black squares represent the individual
    hydrodynamic models, the solid line represents the fit used
    in this work (Equation~\ref{eqn:mdotLx}), the dashed line
    represents the linear analytical prediction.}\label{fig:mdotLx}
\end{figure}
Figure~\ref{fig:hole_mdot} shows
the results for inner hole sources, where the left and right panels show the mass-loss rate as a function of hole
radius and X-ray luminosity respectively.

\item The cumulative mass-loss rates (i.e. the radial profiles
  of the surface mass-loss rates)  for different X-ray luminosities are
  fairly self-similar for primordial discs, and also for inner hole
  sources when one considers the profile as a function of $(R-R_{in})$,
  where R$_{in}$ is the inner hole radius\footnote{Throughout this work we use ${R,\phi,z}$ to refer to
cylindrical co-ordinates and ${r,\theta,\varphi}$ to refer to spherical polar co-ordinates.}. Figure~\ref{fig:mdotR} (left panel) shows the
  normalised cumulative surface mass-loss rates as a function of
  radius for different X-ray luminosities for our primordial disc
  models. It is clear from the figure that higher X-ray luminosities
  produce a broader mass-loss profile; however, this has negligible effect on the global disc evolution and we adopt a mean
  profile, shown by the solid black line in the figure. Figure~\ref{fig:mdotR}
  (right panel) shows
  the  normalised cumulative surface mass loss rates as a function of
  R-R$_{in}$ for the inner hole models. The mean profile adopted for
  the inner hole models is shown again as the solid black line.

\end{enumerate}
\begin{figure*}
\centering
\includegraphics[width=\textwidth]{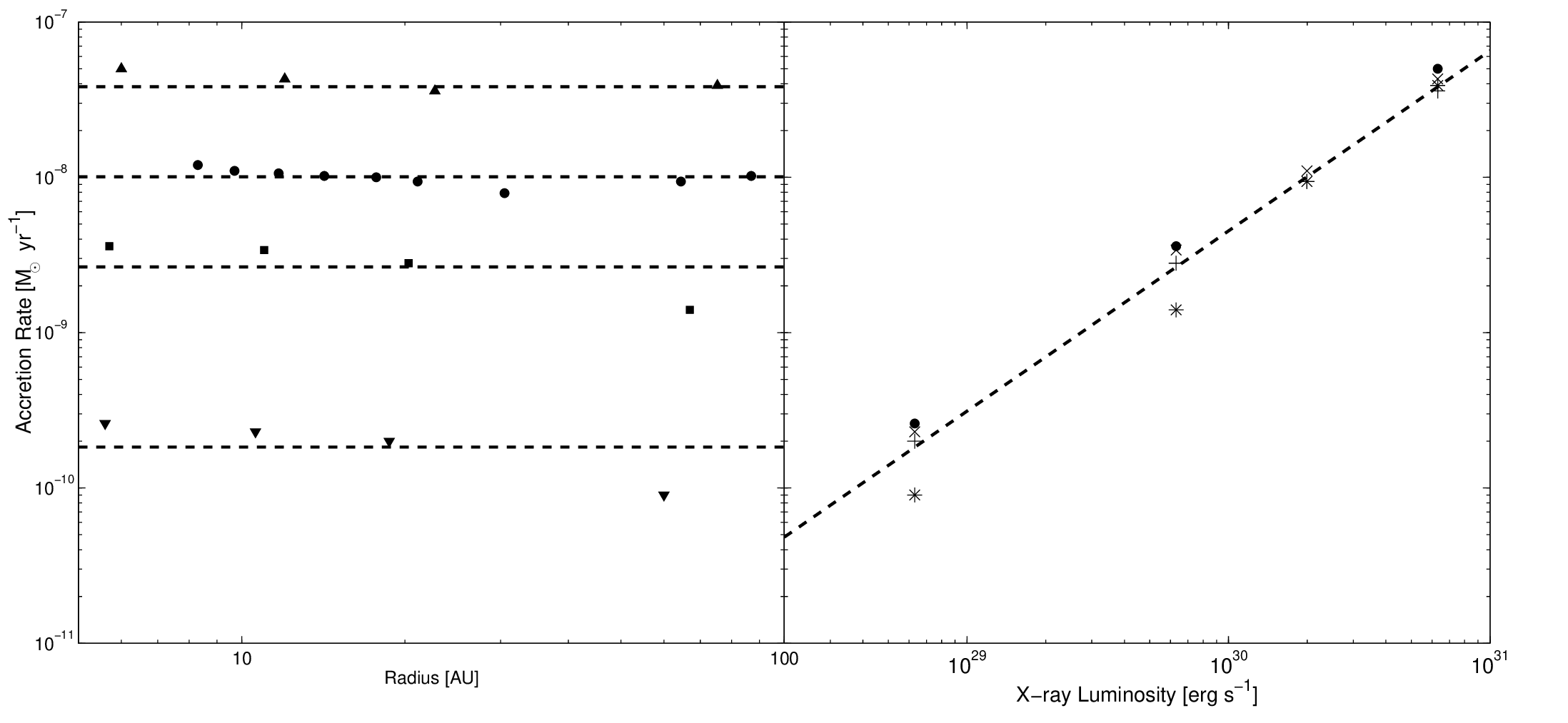}
\caption{Mass-loss rate as a function of
    inner hole radius (left panel), the points show the results from the
    simulations (upright triangles - $L_X=30.8$; circles - results taken from Owen et al. 2010 for $L_X=30.3$; squares - $L_X=29.8$; downward triangles - $L_X=28.8$) and the dashed line shows the fit used throughout this
    work. The right panel shows the mass-loss rate as a function
    of X-ray luminosity ($\bullet$ - $\approx5$AU; $\times$ - $\approx10$AU; $+$ - $\approx20$AU; $*$ - $\approx70$AU)  with the dashed line showing the fit used in
    the viscous evolution.}\label{fig:hole_mdot}
\end{figure*}

\begin{figure*}
\centering
\includegraphics[width=\textwidth]{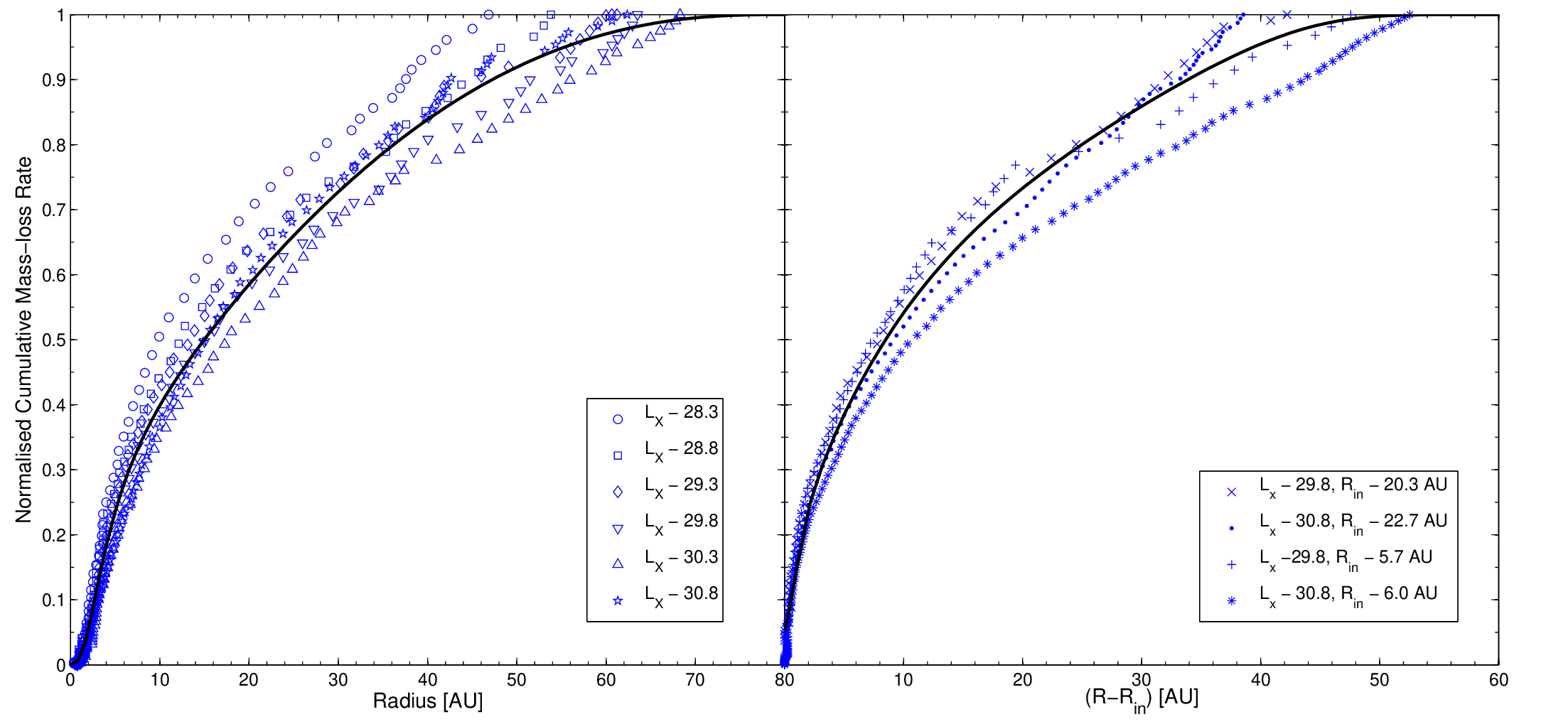}
\caption{Left: radial mass-loss profiles for primordial
    discs. The points represent the results from individual
    hydrodynamic solutions. Right: mass-loss profiles for discs with
    inner holes plotted as a 
    function of $(R-R_{in})$. The points represent a subset of the 21
    hydrodynamic models calculated. All models show the 
    same general profile with higher luminosities and smaller inner
    holes giving rise to slightly broader profiles. In both panels the solid
    black line represent the fits used to both primordial and
    inner hole discs.}\label{fig:mdotR}
\end{figure*}

From the results presented in this section it is immediately apparent that the mass-loss
properties for a star of a given mass are completely controlled by the
X-ray luminosity. We can therefore construct a population synthesis
model that includes only viscosity and XPE (with
appropriate initial conditions, see Section~4.1).

\section{Photoevaporating Viscous Discs}

The evolution of the surface density of a photoevaporating and
viscously evolving disc can be described in one dimension using the
formalism of Lynden-Bell \& Pringle (1974):

\begin{equation}\label{eqn:big}
\frac{\partial \Sigma}{\partial t}=\frac{3}{R}\frac{\partial}{\partial
    R}\left[R^{1/2}\frac{\partial}{\partial R}\left(\Sigma\nu(R)
    R^{1/2}\right)\right]-\dot{\Sigma}_w(R,t)
\end{equation}

\noindent where $\nu$(R) describes the
viscosity term and $\dot{\Sigma}_w(R,t)$ represents the mass-loss due
    to photoevaporation calculated in the previous section. Before
    moving on to discuss our choice of initial conditions and
    viscosity law, it is worthwhile discussing the qualitative
    evolution of a photoevaporating viscously evolving
    disc. Equation~\ref{eqn:big} is perhaps easiest to understand in its
    integral form (i.e $\int_0^\infty\dd R\, 2\pi R\times
    \textrm{Equation}$ \ref{eqn:big}), which tells us:
\begin{equation}
\frac{\partial M_d}{\partial t}=-\dot{M}_*-\dot{M}_w
\label{eqn:bigM}
\end{equation} 
where $M_d$ is the disc mass and $\dot{M}_*$ is the accretion rate
    onto the star due to viscous transport. Equation \ref{eqn:bigM} tells us immediately that
    there are two phases of disc evolution: when $\dot{M}_*>\dot{M}_w$
    the disc evolution is dominated by the viscous transport of
    angular momentum and hence associated accretion rather than the
    removal of gas through photoevaporation.  During this stage the disc
    will behave in a similar manner to a standard viscously evolving disc,
    without photoevaportion especially when $\dot{M}_* \gg \dot{M}_w$. However, when $\dot{M}_w>\dot{M}_*$
    it is now photoevaporation that dominates the evolution of the
    disc, opening a gap in the disc and then finally
    removing the remaining disc material until the disc is
    dispersed.  The accretion rate of a disc evolves on the disc's viscous
    time $t_\nu$ at the outer radius,  which is of order the disc's
    lifetime (Lynden-Bell \& Pringle, 1974; Pringle, 1981; Ruden,
    1993). The transition from a viscously evolving disc to a
    clearing disc occurs when a gap opens in the inner disc and the
    material inside the gap drains
    onto the central star. Since, the time-scale for the inner disc to
    drain is the viscous time at the point the gap opens (which is
    typically much less than the global viscous time of the disc), the
    transition from a viscously accreting disc (i.e. one that would
    observationally be classified as an optically thick accreting primordial disc) to one that is being dispersed, occurs on a time-scale
    much shorter than the disc's current lifetime (Clarke et
    al. 2001; Ruden 2004).

Hence, the evolution of a viscously evolving
    photoevaporating disc can naturally explain the observations of
    discs which appear to evolve under the effect of viscous transport
    alone, but are then rapidly dispersed. A key test for a model of
    viscously evolving photoevaporating discs is to check whether it can self consistently explain
    the observational statistics for {\it both} the
    evolution of primordial discs and those thought to be in the act of
    clearing (i.e. `transition' discs) and the time-scales associated
    with their evolution. As such, any model of disc evolution depends on the choice of viscosity law and the initial
    conditions associated with disc evolution. 

\subsection{Viscosity Law and Initial Conditions}
We now turn our attention to the choice of initial conditions and
viscosity laws that are required to solve Equation~\ref{eqn:big}. We adopt the form $\nu(R) = \nu_0 R$, where $\nu_0 =
\alpha {c_s}^2 / \Omega$, which we always evaluate at 1AU. Such a choice is appealing both
    observationally and theoretically, since it predicts
a surface density scaling as $\Sigma\propto R^{-1}$ which is
supported by observations (e.g. Hartmann et al 1998; Andrews et
    al. 2010) and is consistent
with a constant $\alpha$ disc which is mildly flaring 
i.e. $H/R\propto R^{5/4}$. Furthermore,  we adopt the
zero-time similarity solution of Lynden-Bell \& Pringle (1974) for which
the initial surface density distribution takes the form:
\begin{equation}\label{eqn:start}
  \Sigma(R,0)=\frac{M_d(0)}{2\pi R R_1}\exp(-R/R_1)
\end{equation}
\noindent where M$_d(0)$ and R$_1$ are the initial disc mass and
    a scale
radius describing the exponential taper of the disc's outer region . These initial conditions, together
with the viscosity parameter $\alpha$ and the X-ray luminosity fully
determine the evolution of the disc through Equation~\ref{eqn:big}.

At this point it is useful to construct a `null model'  of
viscous evolution without photoevaporation that when combined with the
    observed X-ray luminosity function can explain the observed
    decline in disc fractions with age. In effect we are asking
    whether there is a universal set of disc viscous parameters which
    can explain the variation in disc lifetime from cluster to cluster
    purely in terms of the observed spread in X-ray luminosity. As discussed above the basic principle of all
photoevaporation models is that discs should evolve viscously, hardly
noticing the effects of photoevaporation, until the mass accretion
rates in the discs have fallen to a value that is comparable to the
photoevaporation rate\footnote{Owen et al (2010) showed that XPE can
  create a gap in a disc only when the accretion rates onto the star are
  approximately an order of magnitude lower than the photoevaporation
  rates.}, at which point  the remaining disc material is rapidly cleared. 

\begin{figure}
\centering
\includegraphics[width=\columnwidth]{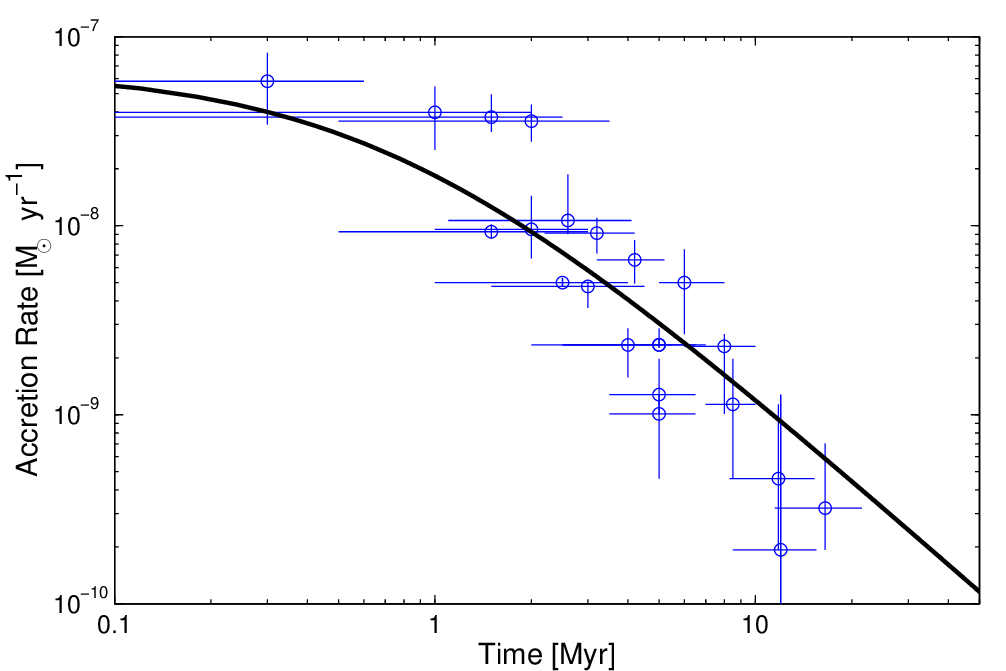}
\caption{Accretion rates as a function of time for the `null' disc
  model. The errors in accretion rate are estimated from the $\sqrt{N}$ errors
    in the disc fractions. The solid line shows a fit of
    Equation~\ref{eqn:mdot} used to determine suitable values of
    $M_d(0)$ and $t_\nu$.}\label{fig:null}
\end{figure}

Viscous evolution alone with $\nu\propto R$ predicts that accretion
    rates evolve as:

\begin{equation}
\dot{M}_{*}(t)=\dot{M}_*(0)\left(1+\frac{t}{t_\nu}\right)^{-3/2}
\label{eqn:mdot}
\end{equation}
where $t_{\nu}$ is the viscous timescale at $R_1$.
This evolution should therefore be observed
in discs before XPE sets in. We can equate the fraction of
disc-bearing pre-main sequence stars at a given time ($f_d$), with the
    fraction of stars in the X-ray luminosity function that have
    luminosities less than a cut-off X-ray luminosity $L_c(f_d)$. In
    order for objects with X-ray luminosities equal to $L_c$ to be
    about to lose their discs, the viscous accretion rate must be
    equal to $\dot{M}_w(L_c)$ at this point. We have performed this simple exercise using the disc
fractions for nearby clusters compiled by Mamajek (2009), the Taurus X-ray luminosity function and the XPE theory
developed above. The result is shown in Figure~\ref{fig:null}, where each point
represents the current accretion rate cut-off in a cluster implied by
XPE. We scale our results on disc fractions assuming an initial close
binary fraction of 14\% by considering that the 0.3Myr old cluster
NGC~2024 (Haisch et al. 2001a), which shows a disc fraction of 86\%, is
too young for any disc to have been destroyed by photoevaporation or
planet formation, but only through binary interactions. The solid line
in the plot represents a suitable fit of Equation~\ref{eqn:mdot} to
    the data, from this fit we can extract an initial accretion rate
    of $\dot{M}_*(0)=5\times10^{-8}$\msunyr and a viscous time of
    $t_\nu=7\times10^{5}$yr. From these two values we can calculate
    an initial disc mass of $M_d(0)=0.07$\msun, which is similar to
    the canonical value (10\% of the stellar
    mass) at which viscous angular momentum
    transport takes over from self-gravity. 

Along with giving us appropriate initial conditions for our disc
    population model, the above also provides a stringent test of the
    hypothesis that X-rays are key to disc evolution and dispersal. If
    the X-rays were not the dominant dispersal mechanism, there is no
    a priori reason to expect a `null' model constructed only using
    knowledge of the X-ray luminosity function and observed disc
    fractions to reproduce a plausible evolution of the accretion
    rates seen in CTTs, both in terms of the time exponent (in
    Equation~\ref{eqn:mdot}) and its initial value. In fact,
    increasing or decreasing the spread about the median of the Taurus
    X-ray luminosity function by a factor of $\sim$5 or greater makes
    a fit of Equation~\ref{eqn:mdot} to the `null' model
    impossible. Although this agreement could be fortuitous, it is
    reassuring that the X-ray luminosity function, disc lifetimes and
    accretion histories are consistent with our XPE hypothesis.  

\begin{figure}
\centering
\includegraphics[width=\columnwidth]{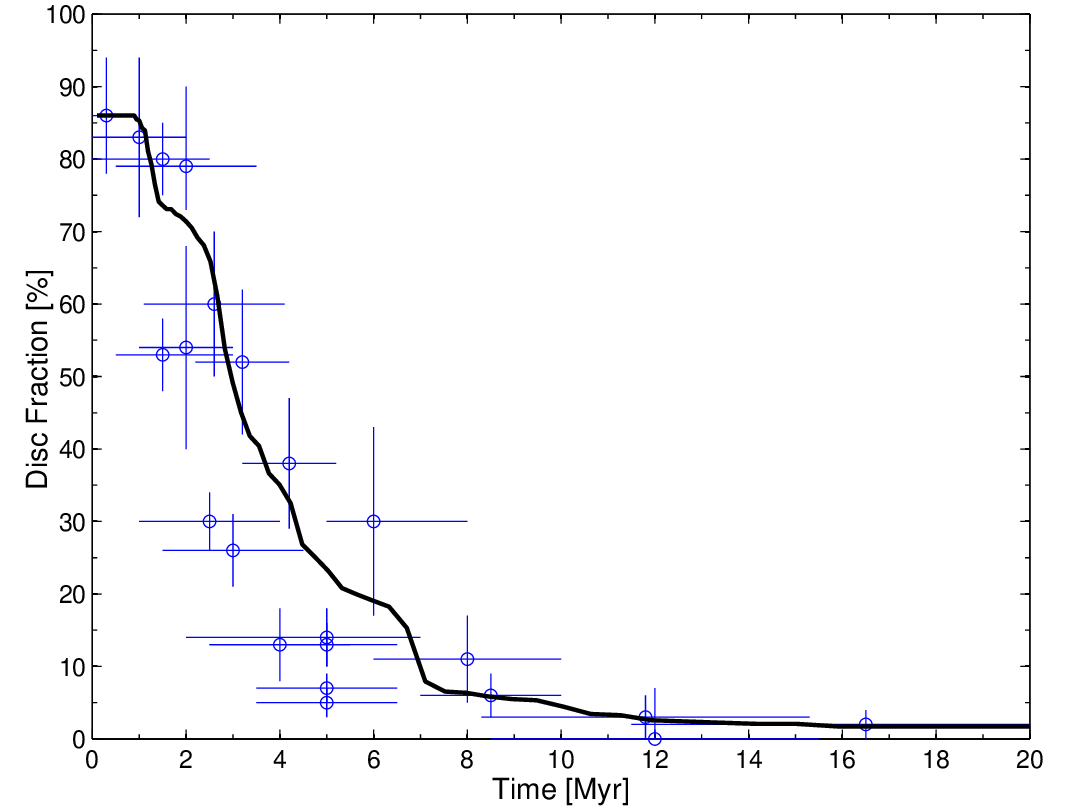}
\caption{Primordial disc fraction as a
    function of time (solid black line) from our XPE population calculated using 500 disc
    models with the mass-loss rate determined by randomly sampling
    the Taurus X-ray luminosity function, all discs evolve from a single set of initial
    conditions. The points are observed disc fractions compiled
    by Mamajek (2009). Model disc fractions have been scaled to
    account for disc destruction by close binary interaction. Note
    that the structure of the black line reflects the structure in the
    X-ray luminosity function.}\label{fig:df}
\end{figure}

Furthermore, in order to uniquely specify the viscous evolution we must
    pick suitable values of $\alpha$ and $R_1$ which in turn specifies $\nu_0$.
While any combination of $R_1$ and $\alpha$ that give the required viscous
    time will reproduce the same `null' viscous model
    (i.e. non-photoevaporating), a disc evolution model that includes
    photoevaporation mildly breaks this degeneracy. By performing a fit (by eye) of a
    viscously evolving photoevaporating disc population to the disc
    fractions used to derive the `null' model we obtain values of
    $\alpha=2.5\times 10^{-3}$ and $R_1=18$AU although due to the
    large scatter in the disc fractions and only mild breaking of the
    $\alpha$, $R_1$ degeneracy, we estimate errors on these values of
    a factor of 2-3. In Figure~\ref{fig:df} we show the obtained fit
    to the disc fractions compiled by Mamajek (2009) for our viscously
    evolving photoevaporating disc model with a single set of initial
    conditions: $M_d(0)=0.07$\msun, $\alpha=2.5\times10^{-3}$ and
    $R_1=18$AU. 

\begin{figure}
\centering
\includegraphics[width=\columnwidth]{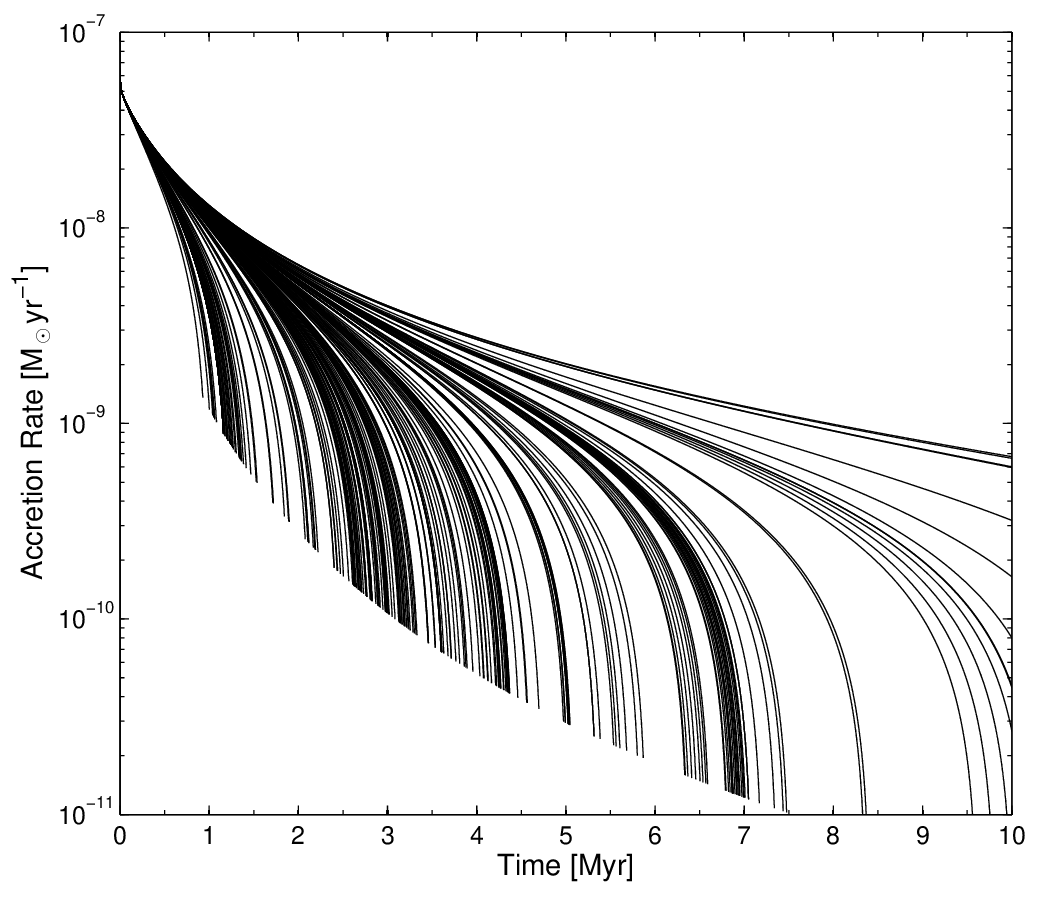}
\caption{Each solid line represents the evolution of accretion rate as
    a function of time for an individual disc model undergoing X-ray
    photoevaporation. All models have the same initial conditions, the
    X-ray luminosity is the only variable.}\label{fig:mdott}
\end{figure}

Finally, Figure~\ref{fig:mdott} shows the predicted evolution of
    accretion rate with time for individual disc models (each line shows
    the evolution of accretion rate with time for one disc model, for a
    given value of the X-ray photoevaporation rate). This clearly
    shows that a spread in accretion rates at late times $>1$Myr does
    not necessary require a spread in initial conditions as some authors
    have suggested is necessary outside the XPE framework
    (e.g. Armitage et al. 2003; Alexander \& Armitage, 2009).
    Observationally there is however, a spread  in accretion rates seen at early times $<$1Myr
    (e.g. Hartmann et al. 1998). This variability cannot be fit with a single
set of initial conditions since photoevaporation has had no time to
    act on the disc. The failure to match the spread in
    accretion rates at early times $<1$Myr may not be surprising,
    since at early times
    the angular momentum transport
    mechanism may be dominated by self-gravity  and accretion may be
    episodic (e.g. Lodato \& Rice, 2005). In reality the zero time point in our models
    corresponds to the point at which the transition from a
    self-gravitating to a viscous disc occurs, as indicated by the
    determination of $M_d(0)=0.1M_*$. This may explain, why our initial accretion rate is significantly lower than some of the accretion rates measured for the youngest objects (Hartmann et al. 1998). 
    
    Furthermore, better agreement for the early time data could easily be obtained
    by assuming a range of initial surface density profiles.
    Indeed our choice of a self-similar
    surface density distribution at zero time has no physical basis
    other than convenience (in fact it would be extremely surprising for discs to
    be born with the surface density distribution of the zero time similarity
    solution of Equation~\ref{eqn:big} given viscous angular momentum
    transport is not a key process during the disc formation stage).
    We emphasise, that any initial surface density distribution with the same
    initial viscous parameters $M_d(0)$, $t_\nu$ and $\alpha$ will tend
    to the same evolution after a few viscous times
    as shown in Lynden-Bell \& Pringle (1974). For this
    reason, we do not attempt to explain disc evolution at
    early times here,  since we are mainly interested in the question
    of disc dispersal, for which only the viscous evolution phase at
    $>1$Myr is relevant. 

\section{Results and Discussion}

We have used the methods and initial conditions derived in the
previous section to construct a population
synthesis model for the evolution of discs dominated by viscosity and
XPE. We have calculated 500 disc models based on a random sampling of the
    Taurus X-ray luminosity function. Our disc evolution models are
    computed by solving Equation~\ref{eqn:big} numerically using the
    method set out in Owen et al. (2010), following the evolution of
    the disc until the disc is cleared to 100AU. At radii larger than 100AU the
    photoevaporation rates are extremely uncertain and we cannot be
    confident in results that continue the
    evolution beyond this radius.  However, for the sake of completeness we will discuss
    the possible qualitative evolution of these remnant discs in Section~5.3. We now turn our attention to
    some specific predictions from these models and, where
possible, compare them with observations.

\subsection{Photoevaporation starved accretion}

As discussed in Section~2, Drake et al. (2009) suggested that coronal X-rays suppress the
accretion flow onto young solar-type stars through the driving of a
photoevaporating wind. This photoevaporation
starved accretion phase can explain the tentative negative correlation
between mass accretion rate and stellar X-ray luminosity reported by Drake et al
(2009). Moreover the reduction in disc lifetime in strong X-ray
    sources can explain the observation that the X-ray luminosities of
    accreting T Tauri stars are systematically lower than those of non-accretors.

\begin{figure}
\centering
\includegraphics[width=\columnwidth]{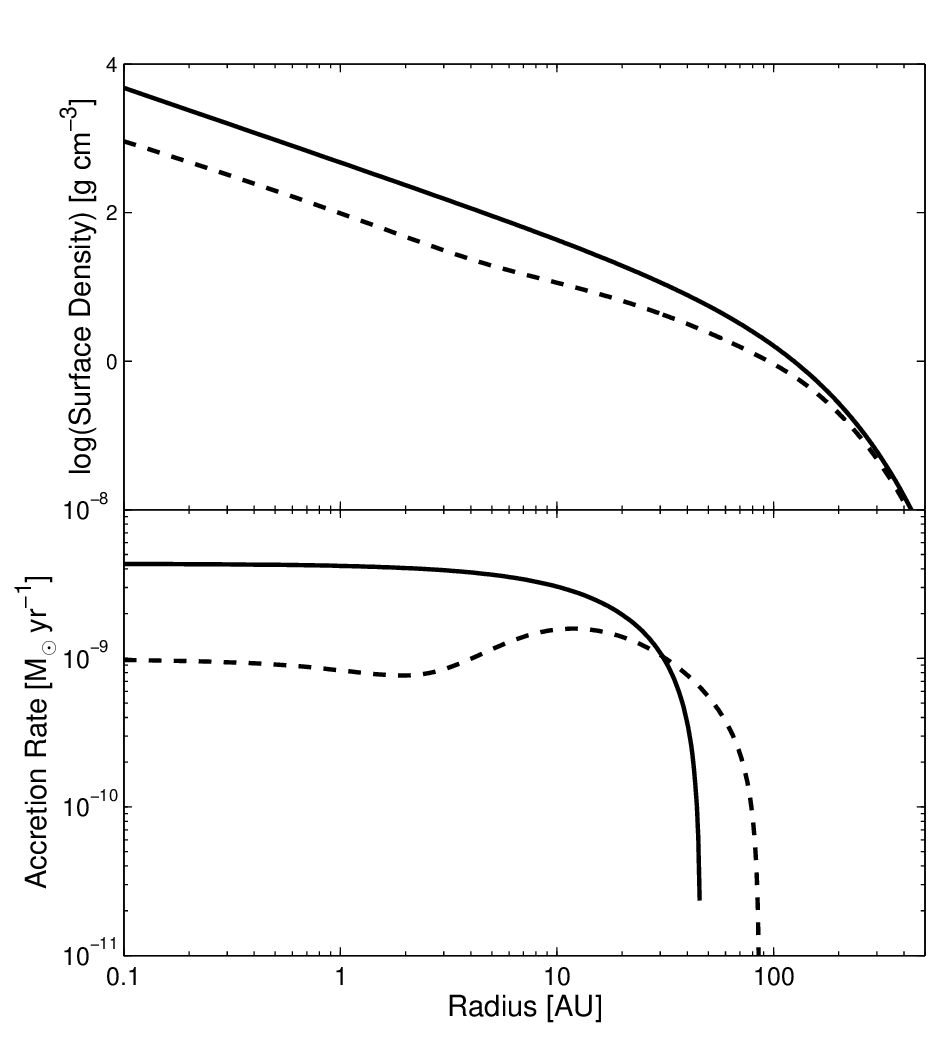}
\caption{Surface density (top) and accretion rate (bottom) radial
  profiles. The solid line represents a
    disc undergoing no photoevaporation and the dashed line represents
    the median disc model $\approx0.5$Myr before a gap opens in the
    disc due to XPE.}\label{fig:psa}
\end{figure}

The XPE models of Owen et al (2010) clearly show that there is indeed a phase
in the disc evolution (before the opening of the gap) where the
effects of this `starving' are apparent in the radial dependence of the accretion
rate. In Figure~\ref{fig:psa} we compare the accretion rate and
surface density profiles of the median disc model 
undergoing XPE, 0.5Myr before the gap opens,  against those of a disc which is only subject to viscous
evolution. Inside $70$ AU the accretion rate drops
before it reaches the star compared to the standard case where the
accretion rate tends to a constant throughout the entire disc. This
    can be compared to the EUV photoevaporation model: in this case
    the mass-loss profile is narrowly peaked between $1-10$AU (for
    solar type stars) and the total mass-loss rate is considerably
    less ($\sim 10^{-10}$\msunyr). This results in a shorter and much
    less pronounced period of `starving' i.e. the disc is only
    affected inside a few AU.  In contrast, the photoevaporation starved
    accretion lasts for $\sim$20-30\% of the disc lifetime in the X-ray model, with
    significant consequences for global disc evolution: i.e. a
    flattening of the surface density profile and a significant drop in the
    accretion rate through the disc.  

\begin{figure}
\centering
\includegraphics[width=\columnwidth]{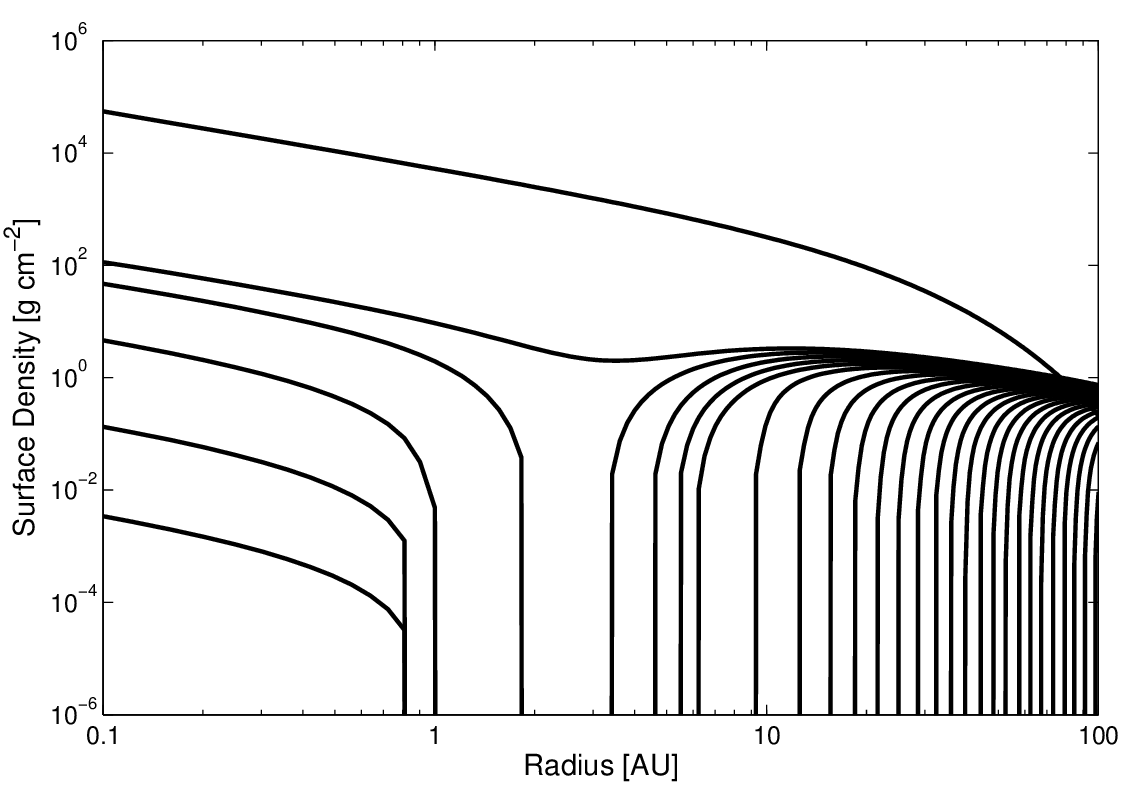}
\caption{The evolution of the disc's surface density during the disc
    clearing phase. The first line shows the zero time surface density
    profile, the next shows the profile at 75\% of the discs lifetime
    ($\sim3.5$Myr) and the remaining lines show the surface density at 1\%
    steps in disc lifetime.}\label{fig:median}
\end{figure}

In Figure~\ref{fig:median} we show the evolution of the surface
    density for the median disc model  (i.e. the disc with the
    median X-ray luminosity of $1.1\times10^{30}$~erg s$^{-1}$ which
    corresponds to  a photoevaporation rate of $7.1\times10^{-9}$~\msunyr) undergoing the stages of
    gap-opening and final clearing.    This shows the drop in
    surface density through the disc between $1-70$AU before the gap
    opens (as shown in Figure~\ref{fig:psa}). Moreover the broad
    photoevaporation profile also causes the steady erosion of the
    disc during the draining of the inner hole, so that the hole
    (though opening at $\sim3$AU) roughly doubles in size during inner
    hole draining. Once the inner disc has completely
    drained there is a rapid clearing of the disc out to
    $10-20$AU, because this region was previously depleted during the
    photoevaporation starved accretion phase. The fast clearing phase slows
    down once the inner hole reaches radii less affected by
    photoevaporation starved accretion. This results in a 
    a total clearing time of roughly $10-20$\%  of the disc lifetime,
    which is consistent with transition disc statistics. This can be
    compared to the EUV  models of
    Alexander et al. (2006b), which resulted in a transition phase approximately
    3\% of the disc's lifetime, and those of Clarke et al. (2001)
    which clear the
    outer disc on time-scales of the order of the disc lifetime.

The negative $\dot{M}$-$L_X$ correlation reported by Drake et al (2009)
is a simple consequence of the fact that discs with higher X-ray
luminosities produce more vigorous winds causing the disc's accretion
rates to be lower than those for discs with a less vigorous
wind. Thus a  negative $\dot{M}$-$L_X$ correlation  is expected for clusters with
a relatively narrow age range. This effect is however counteracted by
the fact that discs with lower X-ray luminosities take longer to evolve,
spending more time at lower accretion rates compared to high
luminosity objects. Therefore, if the X-ray luminosity was compared to
accretion rates for an entire disc population's lifetime, in clusters
    with very large age spreads a positive
correlation might be expected.

\begin{figure*}
\centering
\includegraphics[width=\textwidth]{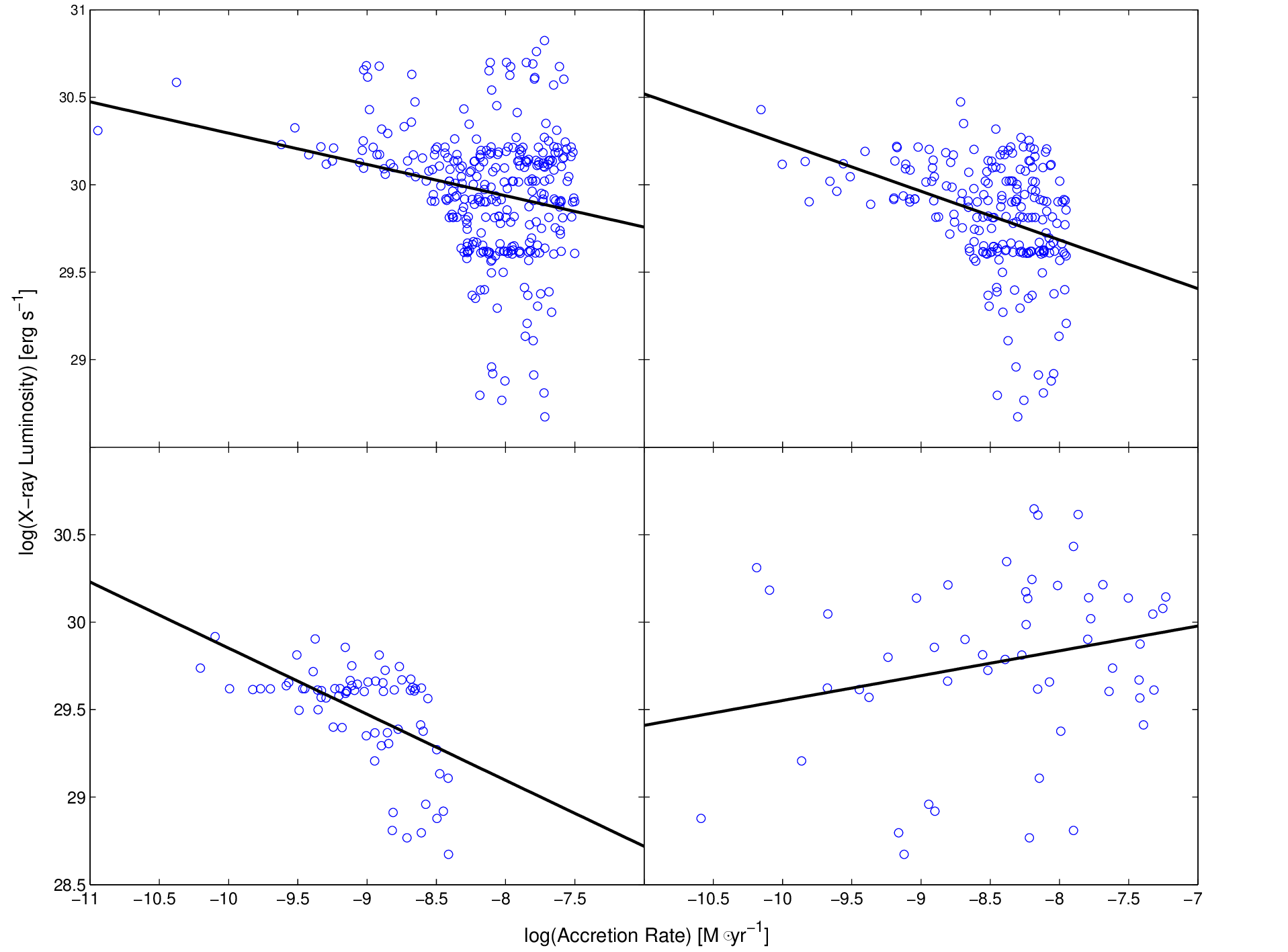}
\caption{Synthetic observations of X-ray luminosity and
    accretion rates of clusters with a uniform age spread, the
    synthetic observations are shown as points while a linear fit to
    the data is shown as the solid line. The top left and right panels
    and the bottom left panels represent young clusters with uniform age
    spreads of 0.5-3.5Myr, 1.5-4.5Myr and 4-8Myr, respectively. The
    bottom right panel represents the disc population observed over
    the entire evolution.}\label{fig:syn_obs} 
\end{figure*}

In Figure~\ref{fig:syn_obs} we show plots of $\dot{M}$-$L_X$ for our
synthetic `cluster' members selected to give clusters with various
narrow age ranges (upper panels and 
lower left panel) along with the a plot of the total disc population
(bottom right panel). As expected, the XPE model
predicts a positive $\dot{M}$-$L_X$ correlation for clusters with
large age spread and a negative correlation for clusters with a narrow
age spread. As a comparison, the age spread in Orion is roughly 2-3Myr
(Haisch et al. 2001b), which explains the observation of a negative correlation by
Drake et al. (2009).  

\begin{figure}
\centering
\includegraphics[width=\columnwidth]{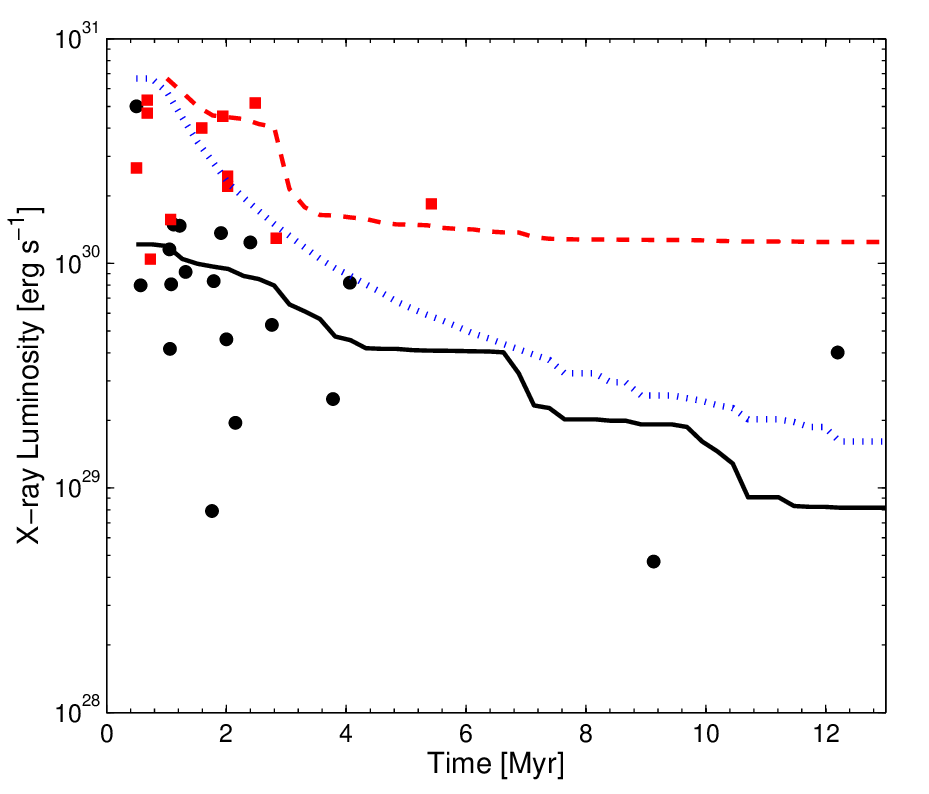}
\caption{Time evolution of the median X-ray luminosity in the synthetic cluster
  for CTTs (solid line) and WTTs (dashed line). The
    dotted line represents the transition X-ray luminosity between a
    CTTs and a WTTs as a function of time. The data points are taken
    from the compilation of G\"{u}del et al. (2007) in the Taurus
    cluster, filled circles refer to CTTs and filled squares
    refer to WTTs. }\label{fig:CTTvWTT}
\end{figure}

The observation of systematically higher X-ray luminosities of non-accreting
TTs (WTTs) compared to accreting TTs (CTTs) can also be explained by
our models in terms of photoevaporation starved
accretion. Figure~\ref{fig:CTTvWTT} 
shows the time evolution of the median  X-ray luminosity of the CTTs
    (solid line) and WTTs (dashed line)
populations for our model compared to the data compiled by G\"{u}del et al. (2007) for
    the Taurus cluster (black circles and red squares). We note the
    good agreement between the model predictions and
    observations. The dotted line represents the critical X-ray
    luminosity as a function of time which in our model separates CTTs
    and WTTs and corresponds to the X-ray luminosity of objects
    which have just opened a gap and have begun the clearing phase. The number of anomalous objects (i.e. CTTs above the
    line and WTTs below the line) is small at ages $>\sim$ 1
    Myr. Given uncertainties in age determinations (particularly at
    $<$ a Myr) the agreement between the observations and predictions
    is very encouraging. 
 
\subsection{The nature of transition discs: accreting and non-accreting}

\begin{figure}
\centering
\includegraphics[width=\columnwidth]{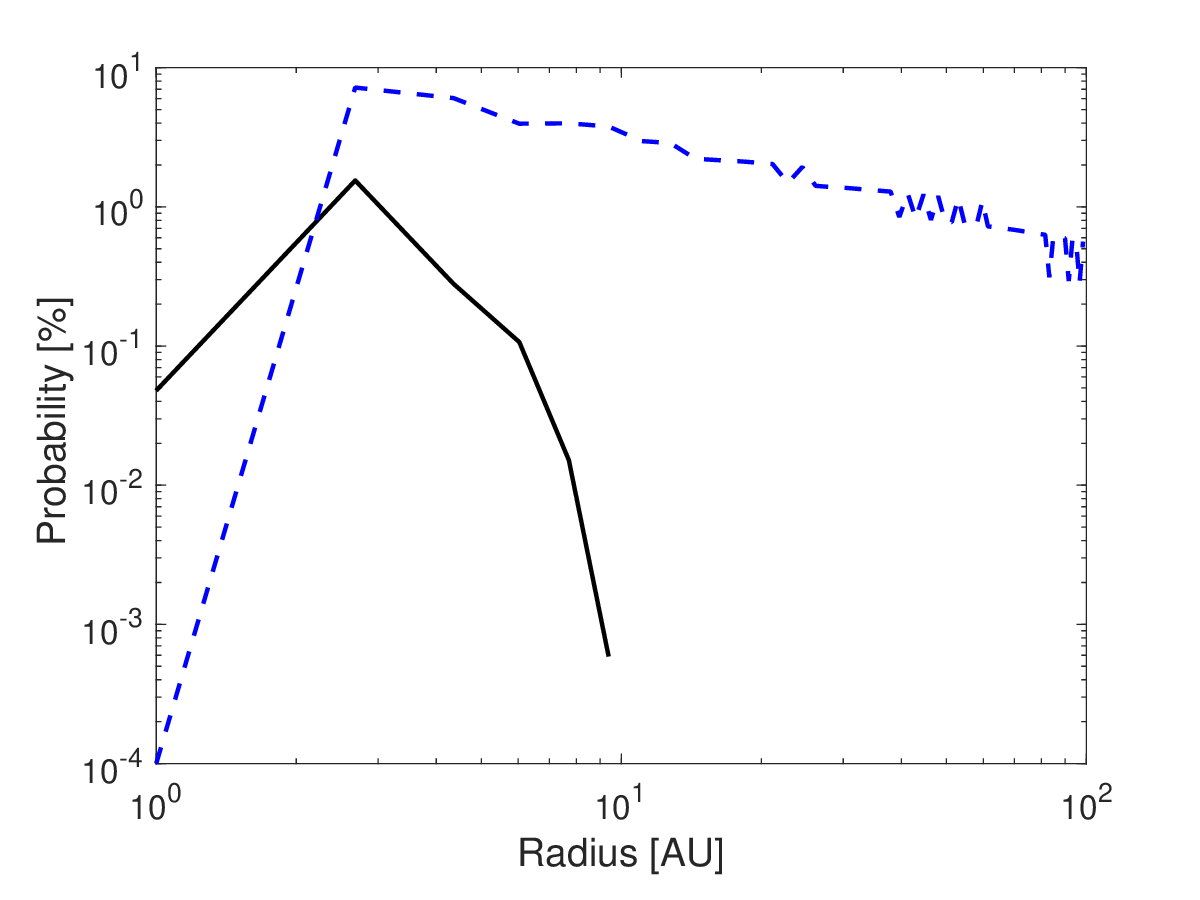}
\caption{Probability distribution function of
    accreting (solid line)  and non-accreting (dashed line) transition
    discs with inner hole radius, they are both scaled so that the integrated sum of both distributions is 100\%.}\label{fig:prob_holes}
\end{figure}

The recent observations of a class of transition (inner hole) discs
with residual gas inside the inner dust radius and with signature of
accretion has prompted some authors to question the viability of
photoevaporation as the formation mechanism for the inner hole in some sources
(e.g. Cieza et al 2010). Previous EUV-driven photoevaporation models
(e.g. Alexander \& Armitage 2009) indeed predicted that at the time of gap opening
the surface density of the gas in the inner disc and the accretion
rates due to the inner disc draining onto the star should be
undetectable. This is however not the case for XPE, mainly due to the
fact that the wind rates can be two orders of magnitude higher than
the EUV-driven rates, meaning that at the time of gap opening the
mass of the draining inner disc and the accretion rate onto the star
of the inner disc material remain detectable for a non-negligible
amount of time. For the disc population generated in this work, 
Figure~\ref{fig:prob_holes}, shows that the accreting inner holes and
    non-accreting inner-holes ($\dot{M}$ $<1\times10^{-11}$\msunyr) are in
    general equally likely out to a radius of $R_{in}\sim5$AU. Clearly as the transition disc is further photoevaporated
and its inner radius moves out the accretion signatures onto the star
become less evident and non-accreting inner holes dominate at radii
larger than 20 AU. The total integrated ratio out to 10AU (the radius probed
    by 24$\mu$m emission around solar-type stars) is found to be $25\%$ accreting and $75\%$
    non-accreting for the entire population. We caution that this is
    {\it not} equivalent to the
    observed fraction of accreting to non-accreting objects in a
    individual cluster, where the cluster age should also be accounted
    for. In young clusters the transition disc population is dominated
    by high X-ray luminosity objects which give rise to a 
    considerably longer accreting inner hole phase. In contrast this
    ratio is much lower in old clusters where the transition
    disc population is dominated by low X-ray luminosity objects that have
    very short accreting inner-hole phases. 

It is perhaps worth noting at this point that the detection of a
`transition' disc observationally is made through observation of the
dust continuum spectral energy distribution (SED). Alexander \& Armitage
(2007) examined the behaviour of dust in a photoevaporating disc,
finding that, under the action of dust drag, the time-scale for
dust grains to drain onto the star is of order 10$^3$yrs, after the gap opens,
approximately two order of magnitudes faster than the gas draining
time-scales. This means that an observer would certainly see a significant drop in opacity in the inner disc immediately after a gap has opened, while the gas will still
linger in the inner dust disc for the duration of its viscous draining
time-scale of $\sim10^5$yrs.

\begin{figure}
\centering
\includegraphics[width=\columnwidth]{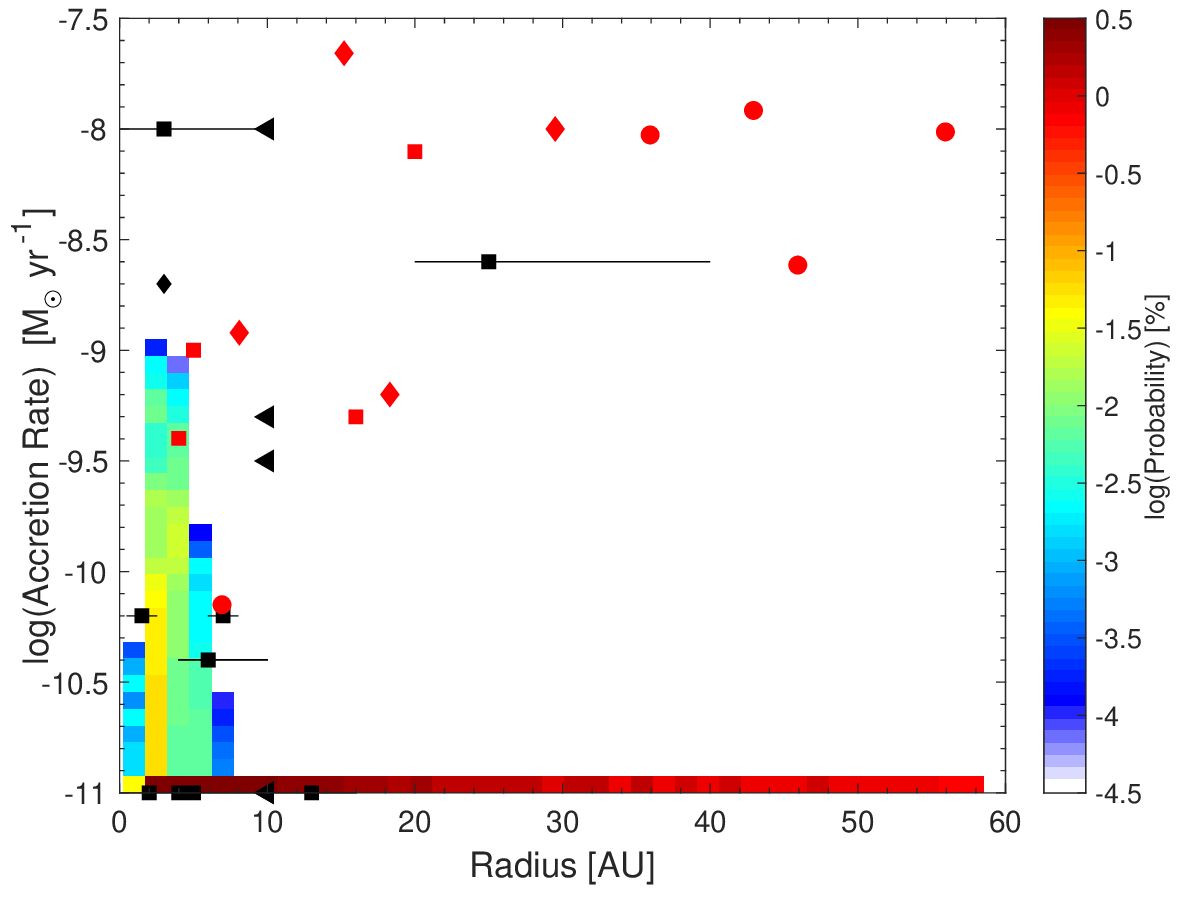}
\caption{Transition disc probability map in the $\dot{M}$-$R_{in}$
  plane. We have applied an observational cut-off of $10^{-11}$\msunyr
  in accretion rate and objects below the cut-off are all shown at the
  bottom of the diagram. We show a representative sample of solar-type objects classified as `transition' discs; data from taken from  Espaillat et al. (2007a,b,2008,2010) - Red Circles, Hughes et al. (2009,2010) - Red Squares, Kim et al. (2009) - Red Diamonds, Calvet et al. (2005) - Black Diamonds, Mer{\'{\i}}n et al. (2010) - Black Squares and Cieza et al. (2010) - Black Triangles, where error-bars are listed they are shown.}\label{fig:innerholes}
\end{figure}

We have used our population synthesis model to investigate the
accretion rate versus inner hole size evolution for transition discs
created by XPE under the assumption of immediate dust clearing at the
time of gap opening. Figure~\ref{fig:innerholes} shows the probability distribution of
the disc models in the $\dot{M}$-$R_{in}$ plane. The symbols
represent a sample of observations of solar-type objects classified as
    `transition' discs by Espaillat et al. (2007a,b,2008,2010) - Red
    Circles, Hughes et al. (2009,2010) - Red Squares, Kim et al. (2009) -
    Red Diamonds, Calvet et al. (2005) - Black Diamonds, Mer{\'{\i}}n et
    al. (2010) - Black Squares and Cieza et al. (2010) - Black
    Triangles (Although Cieza et al. 2010 do not fit for the
    inner-hole radius they list as transitional sources those discs
    that have a deficit of emission in the {\it Spitzer} IRAC bands;
    therefore we conservatively estimate an inner hole radius of
    $<10$AU for all their sources).

   It is
immediately clear from the figure that there is a population of large
inner hole, strongly accreting transition discs that cannot have been
created by XPE. Gap opening by a giant planet or grain growth is
perhaps the most plausible explanation for these objects. However,
there is a significant number of discs with inner holes that
are consistent with an XPE origin. 
    Furthermore we note the lack of observations of non-accreting
    `transition' discs with holes at radii greater than 20AU,
    where our model predicts a significant populations (although
    several non-accreting discs with large holes have been detected in
    different mass ranges e.g. Mer{\'{\i}}n et al. 2010, where the observations probe different radial scales).
The observations are still rather sparse and it is currently not
possible to say whether the observed population of transition discs is a true representation or an artifact of observational selection effects. 

\begin{figure*}
\centering
\includegraphics[width=\textwidth]{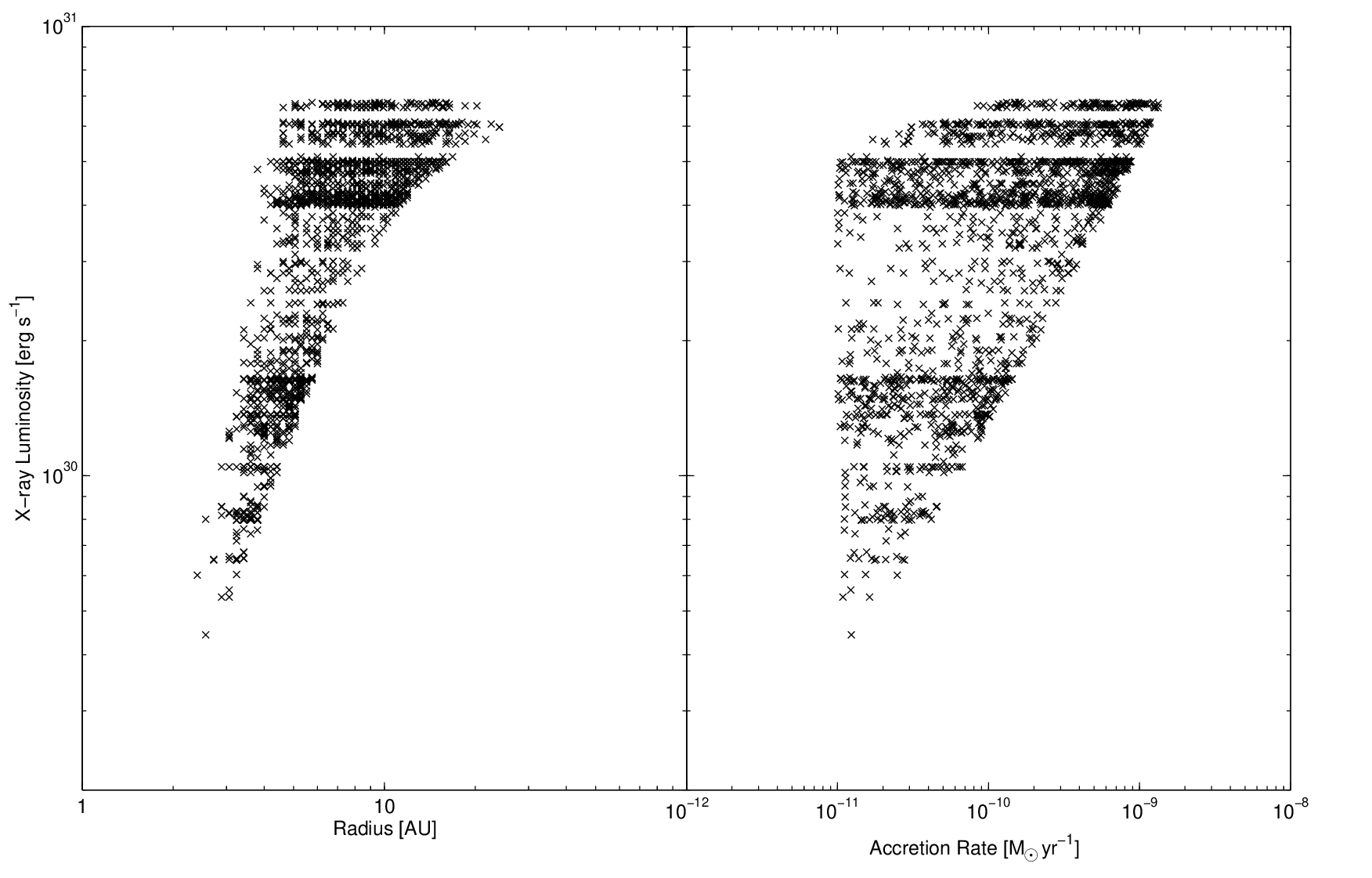}
\caption{Simulated observations of {\it accreting}
    `transition' discs. The left hand panel shows  X-ray luminosity
    plotted against inner-hole radius. The right hand panel shows
    X-ray luminosity plotted against accretion rate where an
    observational cut-off of $10^{-11}$\msunyr has been used.}\label{fig:inner_corr}
\end{figure*}

One obvious consequence of an X-ray photoevaporation mechanism is that
    the properties of transition discs should be correlated with the
    X-ray luminosity, something no other model of photoevaporation or
    `transition' disc origin would predict. In
    Figure~\ref{fig:inner_corr}, we show two such correlations namely
    the inner hole radius (left panel) and accretion rate (right panel) against X-ray
    luminosity, considering only accreting `transition' discs
    (i.e. those with an accretion rate $>1\times
    10^{-11}$\msunyr). The plots have been generated by randomly
    sampling (in time) the accreting  transition phase of each disc
    model several times, and should therefore provide a reasonable
    estimate of both the general form of the correlation plus the
    associated scatter. Clearly, since discs with higher X-ray
    luminosities open gaps earlier and at higher accretion rates a
    strong positive correlation between $L_X$ and $\dot{M}$ would be
    expected and is reproduced by the models as shown. 

Furthermore, we
    also recover a positive correlation between inner hole radius and
    X-ray luminosity for {\it accreting} transition discs as shown in
    the figure. We can understand this by remembering that while the
    inner disc is still draining the inner hole radius is also being
    eroded outwards as shown in Figure~\ref{fig:median}. Since, for
    more X-ray luminous objects the region being eroded is more
    depleted due to photoevaporation starved accretion and the
    magnitude of the mass-loss is higher, the inner-hole radius is
    able to be eroded to larger radius during the accreting
    phase. This explains the positive slope of the right hand extent
    of the symbols in the left panel of Figure~\ref{fig:inner_corr}.  

\subsubsection{Consequences of a different X-ray luminosity function}
 The X-ray luminosity function is a
    crucial input into the XPE model. In this work we have used the Taurus X-ray
    luminosity function as this best represents the quiescent X-ray
    flux the disc sees throughout its lifetime. However, here we
    discuss the consequences of a different X-ray spectrum
    that may be incident on the disc. If the incident spectrum
    the disc sees is systematically harder than the one used (through
    attenuation of the X-ray spectrum by large neutral columns close
    to the star) or softer than the spectrum used, the qualitative
    behaviour of the disc population would remain the same, since one
    can vary the initial condition to fit. Therefore, it is important to
    assess the qualitative changes relating to some of the predictions relating to
    transition discs. As discussed in
    Section~2 it is only the soft X-rays (0.1-1keV) that have any thermal impact,
    then the result of a changing spectrum can just be considered to
    be a change in the soft X-ray luminosity incident on the
    disc. Therefore, overall mass-loss rates can simply be scaled to
    this new harder/softer spectrum. 

In Section~4.1 we argued that any
    X-ray luminosity function with a similar spread can have a `null'
    model constructed to fit the observations of disc fraction. Thus,
    a harder or more attenuated spectrum would result in
    lower-mass loss rates, which would require a lower initial accretion
    rate, compared to a softer spectrum, which would require a higher
    initial accretion rate to explain the observed disc fractions. 
Provided this effect is systematic across all values of $L_X$ (i.e. in doesn't change the spread
    of X-ray luminosities) then  the consequences for the predicted
    `transition' disc population can be considered. For the
    harder/more attenuated spectrum a  lower initial accretion
    rate is required implying a lower initial disc mass, and therefore a smaller
    population of accreting transition discs, with lower accretion
    rates and smaller inner holes. The converse is true for a softer
    spectrum which requires a higher initial accretion rate and hence
    larger initial disc mass, giving rise to a larger population of
    accreting transition discs, with higher accretion rates and larger
    holes.

\subsection{Final clearing of the disc}

Our mass loss rates are only accurate out to $\sim$100~AU, since at this
point other clearing mechanisms (e.g. FUV photoevaporation) may become
dominant. For this reason, we have stopped our viscous evolution models
once the inner hole reaches a size of 100~AU. If we were to
    extrapolate our models to clearing beyond 100AU using Equation~2, we would
find that in about 10\% of cases (sources irradiated by a low X-ray flux)
the final clearing time-scales would exceed 10Myr, resulting in a
population of long-lived discs with large inner holes. These discs may not
survive long if FUV photoevaporation is efficient at large radii as
suggested by Gorti \& Hollenbach (2009). However, given the
uncertainties in the FUV rates, we also consider the case
where these objects survive for a long enough time for their dust
continuum emission to be observed. The spectral energy distribution
(SED) of these cold massive discs should be similar to
that of young debris discs, where a debris disc model is normally considered to be a
    single temperature belt of optically thin dust (thought to be constantly
    replenished by collisions between planetesimals) at a given radius
    from the star (Wyatt 2008). This suggests that some of the
    sources that are currently classified as debris discs may in fact
    be `XPE relics'. 

\begin{figure}
\centering
\includegraphics[width=\columnwidth]{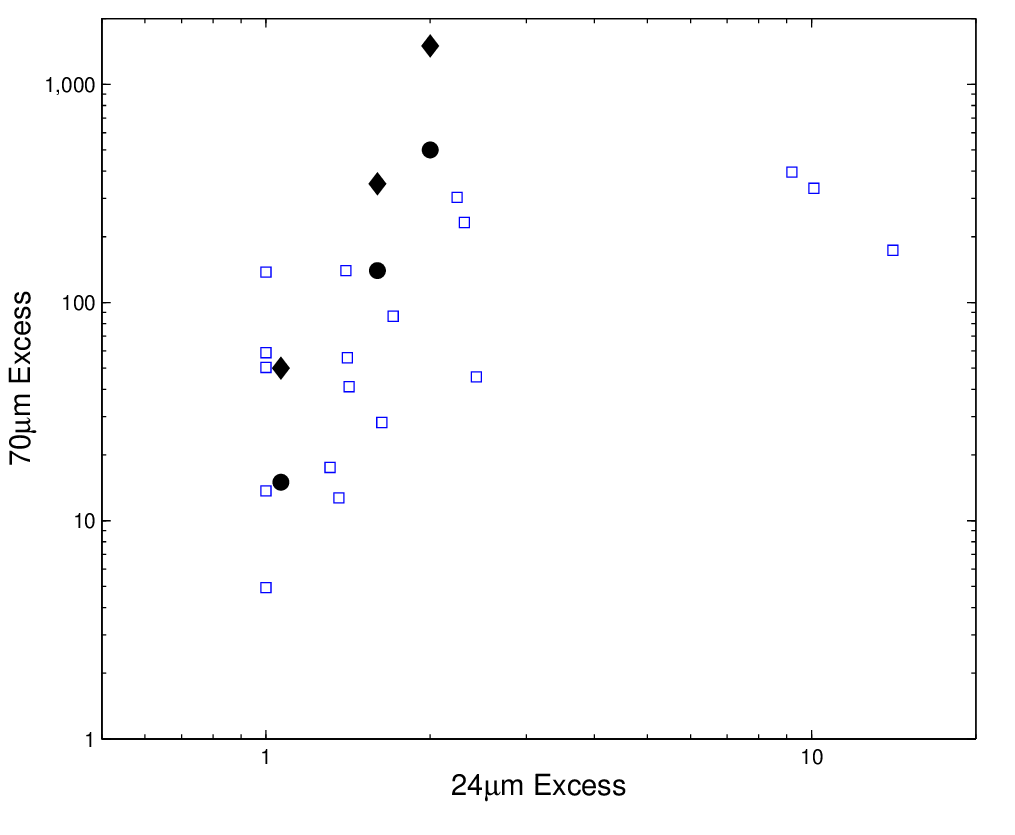}
\caption{Plot of $70\mu$m excess above the photosphere against
    $24\mu$m excess for disc models with a large inner hole (100~AU)
    and low X-ray luminosities and therefore a predicted remaining
    lifetime $>10$Myr The circles show discs close to edge-on and
    increasing $24\mu$m excess represents the evolution from a flat to
    fully mixed dust distribution. Similarly the diamonds show the
    disc models close to face-on.  Also shown as open squares is
    observations of young solar-type stars ($\le100$Myr) with IR excesses classified as debris discs, taken from (Wyatt 2008).}\label{fig:debris}
\end{figure}

We have used the radiative transfer code of Whitney et al (2003a,b) to 
calculate the SED of a typical XPE relic using the standard input disc
    structure and dust properties of the code. The properties of these
    XPE relics were taken from the end point of the lowest X-ray
    luminosity (and hence the longest lived) model with an inner
    radius of 100 AU, outer e-folding radius of 310AU and a mass of
    $7\times10^{-3}$\msun  in gas (and a dust to gas mass ratio of
    0.01). While such a disc is likely to be settled in the dust, we
    computed three models: flat dust distribution, fully mixed dust
    distribution  and a disc where
    $H_{dust}(R)=0.1H_{gas}(R)$. Figure~\ref{fig:debris} shows a plot
    of the fractional excess of the disc compared with the stellar
    photosphere at 24$\mu$m and $70\mu$m, where we 
compare our models with observations of objects classified as young debris
discs around solar-type stars (Wyatt 2008). It is clear from this
figure that XPE relics with a degree of dust settling share the same space in the excess-excess plot
as sources currently classified as young debris discs. The fully-mixed
    discs show a $70\mu$m excess which is probably too large to be
    classified as a debris disc, although it is extremely unlikely that any disc could survive for $>10$Myr without undergoing  dust settling.    
We also note that the predicted 850$\mu$m emission from the 
evolved XPE relics falls below the current detection limits at 50 pc (Andrews \&
Williams 2005), and thus these relics would not contradict previous sub-mm
observations of WTTs which show that most are devoid of emission out to 500 AU (Duvert et al 2000). Only ALMA will be able to
separate these large massive discs from canonical debris discs, and
hence confirm or dismiss the existence of this proposed class of
objects, placing constraints on the role of FUV photoevaporation at large radius. 
\section{Conclusions}

We have used radiation-hydrodynamic calculations of X-ray photoevaporated
discs coupled to a viscous evolution model to construct a population
synthesis model, with which we have studied the physical properties of primordial and
transition discs.  The initial conditions and viscosity law are constrained using recent
observations of disc fractions in nearby clusters (Mamajek 2009). We require a
viscosity coefficient of approximately $\alpha = 2.5\times10^{-3}$ to match the
observational constraints.

Our main conclusions can be briefly summarised as follows:

\begin{enumerate}
\item X-rays play a major role in the evolution and dispersal of discs
  around solar-type stars, driving vigorous photoevaporative winds whose rates
 scale linearly with the X-ray luminosity, which are in the range of
 observed accretion rates for T-Tauri stars. 
\item We have constructed a `null' accretion disc model using only
    knowledge of the observed disc fractions and X-ray luminosity functions
    under the assumption that discs are dispersed through X-ray
    photoevaporation. This `null' model shows very good agreement with
    observed accretion rates in YSOs as well as their evolution with
    time, providing further independent confirmation of the viability
    of X-ray photoevaporation as a dominant dispersal mechanism. 
\item X-rays suppress accretion by preventing accreting material from
  reaching the star, since this material is removed through photoevaporation. This `photoevaporation starved accretion' (Drake et
  al 2009) produces a negative correlation between X-ray luminosity
  and accretion rate for clusters with a relatively narrow age
  spread, in agreement with the observational correlation reported by Drake et al.
  (2009) in the Orion data. 
\item Our models successfully reproduce the observation that
WTTs are systematically brighter in X-rays than CTTs. Whereas this
has previously been interpreted as modification of X-ray emission or
detectability by the presence of discs, our models support Drake et
    al. (2009) in reversing the causal
link  - i.e.  that disc lifetimes are regulated by XPE and hence
discless stars are on average those with higher $L_X$.

\item A large fraction ($>\sim 50$\%) of observed transition discs can
    be easily explained by X-ray photoevaporation. There is however a
  population of strongly accreting transition discs with large inner holes $>20$AU that lies outside
  the $\dot{M}-R_{in}$ region predicted by our models, suggesting that
  alternative mechanisms are responsible for their inner hole
  (e.g. binary interaction, grain-growth and/or planet formation). 
\item A fraction of currently observed objects classified as young
    debris discs (on the basis of their excesses at 24$\mu$m and
    $70\mu$m) may in
  fact be the relics of X-ray photoevapored discs, which are predicted
  to be long-lived ($>$10 Myr) for low X-ray luminosity sources.  Future
  mass determination with {\it ALMA} are necessary to shed light onto
  the nature of these objects. 
\end{enumerate}

\section*{ACKNOWLEDGMENTS}
We are grateful to the anonymous referee who helped improve the
    clarity of this work.
The authors would like to thank Thomas Preibisch and Manuel G\"{u}del
    for providing their data on X-ray luminosities and helpful
    discussions regarding X-ray luminosity functions. We also thank Jeremy
    Drake for insightful discussions on photoevaporation starved
    accretion, X-ray observations and the X-ray luminosity
    function. We would also like to thank Mark Wyatt and Mark Booth
    for helpful discussions regarding debris disc observations and to
    Mark Wyatt for providing the sample of debris disc observations
    used in this work. JEO acknowledges support of a STFC PhD studentship and is indebted to
    the University of Exeter Astrophysics Department for hospitality during the completion of
    this work. BE is supported by a Science and Technology Facility Council Advanced Fellowship.

\label{lastpage}

\end{document}